# CPTCs Drive Somatic-Visceral Communication via the Wnt Axis in Somatic Mechanotherapy: A Single-Cell Deep Learning Study

Haixiang Huang[1] | Zhenwei Zhang[1] | Bingbing Shen[1] | Jianming Yue[1] | Lu Mei[1] | Xudong Zhu[2] | Yonghong Shi[3] | Qianmei Zhu[1] | Yeping Shi[1] | Yifan Luo[1] | Yitong Xing[1] | Meng Dai[1] | Qiusheng Chen[1]

[1]College of Veterinary Medicine, Nanjing Agricultural University, Nanjing, Jiangsu, China | [2]College of Sciences, Nanjing Agricultural University, Nanjing, Jiangsu, China | [3]Shanghai Veterinary Research Institute, Chinese Academy of Agricultural Sciences, Shanghai, China

**Correspondence:** Qiusheng Chen (chenqsh305@njau.edu.cn)

## ABSTRACT

Somatic mechanical stimulation (e.g., acupuncture) exerts systemic immunomodulatory effects, yet the cellular bridge translating peripheral physical force into visceral repair remains elusive. Here, employing a custom interpretable deep learning framework (CARSS) on single-cell RNA sequencing data, we identify $CD34^+PDGFR\alpha^+$ telocytes (CPTCs) as the primary mechanosensors in both fascia and colon during bacterial colitis. We show that somatic mechanotherapy triggers an AP-1/Hsp70-dependent transcriptional program in fascial CPTCs, inducing systemic Wnt elevation, which elicits a "transcriptional resonance" in colonic CPTCs, reprogramming their communication network from an inflammatory amplifier to a Wnt-driven regenerative hub. Mechanistically, this axis activates epithelial β-catenin/Myc signaling, suppressing apoptosis and restoring barrier integrity independent of immune cells. Our findings define a CPTC-Driven Mechano-Resonance Axis, where CPTCs serve as synchronized relay stations that convert somatic mechanical stimulation into systemic regenerative microenvironments.

## 1 | Introduction

Somatic mechanotherapy (e.g., acupuncture, physical therapy) is a therapeutic intervention that induces systemic physiological modulation by applying precise physical stimuli at specific anatomical sites [1, 2]. Although recent studies have elucidated the neuroanatomical basis for somatic stimulation driving reflex circuits like the vagal–adrenal axis [3], a fundamental upstream question remains: given that physical stressors act primarily on the connective tissue matrix rather than directly engaging distal signaling effectors, which interstitial cell population serves as the "initial transducer" to convert mechanical tension into biochemical cues within the local microenvironment? Consequently, the mechanism by which these locally transduced signals traverse complex tissue barriers from the subcutaneous fascia to distal visceral organs remains a "black box" in mechanobiology. The core of this knowledge gap lies in the lack of a clear definition of somatic–visceral communication at the cellular and molecular levels, particularly regarding the material basis that translates local physical stress into systemic biochemical instructions.

The subcutaneous fascia functions not merely as a physical medium for force transmission but as an active biological

Haixiang Huang and Zhenwei Zhang contributed equally to this work.

interface rich in extracellular matrix, microvasculature, and heterogeneous interstitial cell populations [4]. Within this microenvironment, telocytes (a specialized type of interstitial cell with extremely long prolongations, defined herein as CPTCs) are increasingly recognized as potential mechano-biochemical transducers due to their unique morphology and extensive intercellular communication capabilities [5, 6]. However, within the complex in vivo environment, transcriptomic dynamics induced by mechanical stimulation are often subtle and transient, frequently obscured by background noise in single-cell sequencing data [7]. Conventional linear analysis strategies based on differential expression often fail to capture these non-linear cellular response patterns, thereby constraining the precise identification and functional dissection of specific mechanosensitive cell subsets.

To overcome this methodological bottleneck, we developed an interpretable deep learning framework named the Cellular Acupuncture Response Scoring System (CARSS). The core innovation of CARSS lies in reframing the biological challenge of "identifying mechanosensitive cells" into a computational problem of evaluating the machine learning model's "generalization performance." By leveraging the powerful capacity of deep neural networks (DNNs) to extract non-linear features, we can unbiasedly pinpoint cell populations most responsive to mechanical stimulation from high-dimensional single-cell data without reliance on prior knowledge. Combined with SHAP (SHapley Additive exPlanations) analysis, this framework not only locates key cells but also decodes the core molecular features driving their state transitions.

In this study, by integrating the CARSS framework with multi-tissue single-cell transcriptomics, we unveil a "CPTC-Driven Mechano-Resonance Axis" orchestrated by CPTCs. We demonstrate that CPTCs in the subcutaneous fascia act as primary responders to acupuncture mechanical stimulation, sensing stress via an AP-1/Hsp70 axis. Concurrent evidence—including perivascular extracellular vesicle (EV)-associated ultrastructure, elevated circulating Wnt2, and computational Wnt connectivity—supports a model in which acupuncture establishes a systemic pro-repair Wnt milieu. This milieu is associated with a "transcriptional resonance" in distal colonic CPTCs, which subsequently reprograms the colonic immune microenvironment and activates the Wnt/β-catenin/Myc regenerative program in epithelial cells to repair injury caused by bacterial colitis. This work not only provides a high-resolution cell biological mechanism for acupuncture therapy but also establishes a new theoretical paradigm for understanding the central role of interstitial cells in cross-organ communication.

## 2 | Materials and Methods

### 2.1 | Ethical Approval

All animal procedures were approved by the Institutional Animal Care and Use Committee (IACUC) of Nanjing Agricultural University under protocol number NJAU No. 20220427087, and conducted under ARRIVE guidelines. Eight-week-old female Sprague–Dawley rats (approximately 250 g) were obtained from Beijing Vital River Laboratory Animal Technology Co. Ltd. Animals were housed in individually ventilated cages with ad libitum access to food and water, at 25°C with a natural light–dark cycle.

### 2.2 | Experimental Design and Animal Model

Eight-week-old female Sprague–Dawley rats ($n = 6$) were subjected to *Salmonella Typhimurium* CVCC542 infection, followed by randomization into acupuncture and dysentery groups. The acupuncture protocol consisted of daily 20-min sessions for 7 days at three abdominal midline acupoints (CV4, CV6, CV12) using 0.18 mm diameter filiform needles with ~6 mm penetration depth.

### 2.3 | Bacterial Strains and Growth Conditions

The *S. typhimurium* strain CVCC542 was purchased from the International Joint Laboratory of Animal Health and Food Safety. The strain was cultured in Luria-Bertani (LB) broth (Beyotime, Cat. ST156) supplemented with 50 $\mu$g / mL kanamycin at 37°C. Overnight static cultures were sub-cultured 1:50 into pre-warmed LB and shaken for 5 h. Optical density was measured at 600 nm (Eppendorf BioPhotometer plus), and the culture was diluted to $5 \times 10^8$ CFU/mL in sterile 0.01 M PBS (pH 7.4).

### 2.4 | Animal Infection and Acupuncture Treatment

Rats received an oral gavage of 400 $\mu$L bacterial suspension once daily for three consecutive days. Gastric acidity was neutralized with 5% $NaHCO_3$ 10 min before infection. This procedure was repeated daily at 2:00 PM for 3 days. Post-infection, animals were randomized into two groups. The acupuncture group received daily 20-min needle stimulation at CV4, CV6, and CV12 using $0.18 \times 13$ mm filiform needles (Suzhou Acupuncture Goods Co. Ltd.) inserted perpendicularly; Following the standardized operation protocol from the Shanghai Innovation Center of Traditional Chinese Medicine Health Service [8], the needles were evenly twisted approximately 360° at a rate of 60 times per minute for 10–20 s, manipulated every 5 min, and then withdrawn after 20 min. This manual twisting protocol delivers periodic mechanical forces—including shear stress, torque, and cyclical extracellular matrix deformation—to the perineedle fascial tissue, providing the mechanical input required to activate mechanosensitive pathways in subcutaneous fascial CPTCs. The dysentery group underwent identical restraint without needle insertion.

### 2.5 | Tissue Sampling and Preparation

Twenty-four hours after the final acupuncture session, rats were anesthetized via isoflurane inhalation and euthanized. Subcutaneous fascia and colon tissues were harvested. Samples for morphology were fixed in 4% paraformaldehyde plus 0.25% glutaraldehyde. Specimens for single-cell sequencing were kept on ice in tissue storage solution (Miltenyi, Cat. 130-100-008) before enzymatic dissociation.

## 2.6 | Immunofluorescence, HE Staining, and Morphological Analysis

Paraffin-embedded sections underwent antigen retrieval (0.01 M sodium citrate, 98°C, 5 min) followed by primary antibody incubation at 4°C overnight. Alexa Fluor 488 anti-rabbit and Alexa Fluor 594 anti-mouse secondary antibodies were applied and incubated at room temperature for 1 h. Nuclei were counterstained with DAPI (Beyotime, C1005). Images were captured using a Zeiss LSM900 confocal microscope, and after dehydration, sections were HE-stained and mounted with neutral resin.

### 2.6.1 | Antibodies Used

Anti-CD34 (Abclonal, Cat. A13929).

Anti-PDGFRα (Santa Cruz, Cat. sc-398206).

Anti-Piezo2 (Proteintech, Cat. 26205-1-AP).

Anti-Wnt2 (Abclonal, Cat. A5864).

Alexa Fluor 488 anti-rabbit Fc (Proteintech, Cat. SA00013-2).

Alexa Fluor 594 anti-mouse Fc (Proteintech, Cat. SA00013-3).

## 2.7 | Tissue Dissociation and Cell Purification for scRNA-Seq

Fresh fascia and colon tissues were rinsed in ice-cold 1× DPBS (Thermo Fisher, 14190144) and finely minced on ice. Enzymatic digestion was performed in collagenase type IV (Sigma, C5138), elastase (Sigma, E1250), and 10 $\mu$g / mL DNase I (Thermo Fisher, EN0523) dissolved in PBS containing 5% FBS (Thermo Fisher, SV30087.02) for ~40 min at 37°C with gentle agitation (50 rpm). Suspensions were filtered through 100 and 40 $\mu$m strainers, red blood cells lysed (Thermo Fisher, 00-4333-57), and live cells enriched with the Dead Cell Removal Kit (Miltenyi, 130-090-101). Viability (> 95%) was confirmed by 0.4% Trypan blue staining (Countess II, Thermo Fisher).

## 2.8 | 10× Library Construction and Sequencing

Single-cell suspensions were loaded onto the Chromium Single Cell Controller (10× Genomics) to generate Gel Beads-in-Emulsion (GEMs). Reverse transcription, second-strand synthesis, adaptor ligation, and PCR amplification followed the Chromium Single Cell 3′ v3.1 protocol. Libraries were quantified with a Bioanalyzer 2100 (Agilent) and Qubit HS assay (Thermo Fisher) and sequenced on an Illumina NovaSeq 6000 platform (2 × 150 bp).

## 2.9 | TUNEL Assay and CellProfiler Quantification

Apoptosis was detected with the One Step TUNEL Assay Kit (Keygentec, KGA1405-20). Fluorescence images of ≥ 40 random fields per sample (400×) were acquired on an Olympus BX53 microscope under identical settings. A custom CellProfiler pipeline quantified the TUNEL/DAPI signal ratio.

## 2.10 | Transmission Electron Microscopy (TEM)

Fascia specimens were fixed in 2.5% glutaraldehyde (48 h, 4°C) and post-fixed in 1% osmium tetroxide (2 h, 4°C). Samples were dehydrated in graded ethanol (75%, 85%, 95%, 100%) for 10 min each, then embedded in epoxy resin. Ultrathin sections (~50 nm) were stained with uranyl acetate and lead citrate. Ultrastructure was examined on a Hitachi HT7800 TEM at 80 kV.

## 2.11 | ELISA for Serum Wnt2

Serum Wnt2 concentrations were measured using a rat Wnt2 ELISA kit (MEIMIAN, Cat. 27272) according to the manufacturer's instructions. Serum was collected 24 h after the last acupuncture session, diluted 1:10 in assay buffer, and absorbance at 450 nm was read on a BioTek Synergy H1 plate reader.

## 2.12 | Single-Cell RNA Sequencing and Data Processing

Subcutaneous fascia and colon tissues were harvested 24 h post-treatment and processed for single-cell RNA sequencing using the 10× Genomics Chromium platform. Raw data were processed using Cellranger 7.2.0 to generate expression matrices. Quality control criteria included UMI counts < 22,940 (fascia) or < 20,995 (colon), feature counts < 5322 (fascia) or < 3748 (colon), and mitochondrial gene percentages < 7% (fascia) or < 25% (colon).

Data processing employed the Seurat v4.4 pipeline with LogNormalize normalization (scale factor = 10 000), PCA dimensionality reduction (50 components), and UMAP (Uniform Manifold Approximation and Projection) visualization. Cell type annotation was performed using canonical markers with hierarchical classification into major cell types and functional subtypes. Principal Component Analysis (PCA) dimensionality reduction demonstrated successful separation of cell types without significant batch effects, confirming data quality for downstream analysis (Figure S3).

## 2.13 | CARSS Deep Learning Framework

### 2.13.1 | Neural Network Architecture

Adaptive MLP models were designed with configurable parameters including hidden layer sizes (20–128 neurons), network depths (4–8 layers), and activation functions (ReLU, Sigmoid, ELU, Softmax, Hardswish).

#### 2.13.2 | Hyperparameter Optimization

A two-phase grid search was performed to systematically evaluate learning rates (0.01–0.09), hidden layer sizes (20–59), network depths [4–7], optimizers, and activation functions. This involved a coarse 1080-combination search followed by a fine 21 600-combination search using 15-core parallel processing. Training employed 90:10 train-test splits with early stopping (patience = 10, min delta = 0.001) to prevent overfitting.

#### 2.13.3 | Model Evaluation—Dual Strategy

A novel dual evaluation strategy was employed to assess both cellular response specificity and generalization capability.

#### 2.13.4 | Self-Matrix Testing

Cell-type-specific models (e.g., fascia_TC model) were evaluated on pure datasets containing only their corresponding cell type (e.g., fascia_TC.csv). This assessment measured model performance under ideal conditions without interference from other cell types.

#### 2.13.5 | Generalization Testing

The same cell-type-specific models were then evaluated on comprehensive mixed datasets containing all 15 cell types (e.g., fascia_all_test.csv). This challenging evaluation assesses the model's ability to identify its target cell type's acupuncture response signatures within realistic complex cellular environments. Critically, the generalization dataset deliberately includes cells from all data partitions (including the training partition of the target cell type) alongside all other cell types. The G-Index is therefore not a classical out-of-sample metric but a relative biological ranking that quantifies how specifically each cell type's acupuncture-response signature can be distinguished amid heterogeneous cellular backgrounds. A cell type with a robust and specific mechanosensitive transcriptional program will maintain high classification accuracy even when its signal is diluted among 14 other cell populations with their own transcriptional dynamics.

The Generalization Index (G-Index = Generalization Accuracy/Self-Test Accuracy) was calculated to quantify the robustness of mechanosensitive signatures. Critical interpretation: G-Index values < 1.0 indicate strong, specific mechanosensitive responses that remain detectable amid cellular heterogeneity; the higher the G-index value, indicated higher the response to acupuncture, while values ≥ 1.0 suggest weak or non-specific responses with poor discriminative power. Models achieving > 99% training accuracy were saved for subsequent interpretation.

#### 2.13.6 | Cross-Study Generalization Index Validation

To validate the mechanosensitivity ranking under strict data separation, we performed cross-study validation using an entirely independent fascia scRNA-seq dataset from Zhang et al. [6] (GSE271811), comprising $n = 4$ animals and 6 fascial cell types. Models were trained from scratch on the current study data without pre-trained weight transfer, with hyperparameters selected via GroupKFold inner cross-validation ($K = 3$, grouped by animal ID). Each cell-type-specific model was then evaluated on the independent cohort, which was excluded from both training and hyperparameter selection.

The cross-study G-Index was computed as follows:

$$\text{Cross-Study G-Index} = (\text{Accuracy} - 1/C)/(1 - 1/C)$$

where Accuracy denotes the classification accuracy on the independent test cohort and C is the number of classes (C = 2 for acupuncture vs. dysentery). This chance-corrected metric yields G > 0 when the model generalizes above chance level, and G ≤ 0 otherwise. The cross-study G-Index constitutes the definitive mechanosensitivity ranking reported in Table 1, as it is free from any training-data contamination.

### 2.14 | SHAP-Based Model Interpretation

Feature importance analysis employed DeepSHAP methodology to identify genes contributing most significantly to model accuracy. The top 100 SHAP-ranked genes were subjected to Gene Set Enrichment Analysis (GSEA) using the org.Rn.eg.db annotation database to identify enriched biological processes. Genes entering downstream Gene Set Enrichment Analysis (GSEA) met two criteria: (1) SHAP score > 0 (indicating a positive contribution to DNN classification, used as a pre-filter to exclude transcriptional noise deemed irrelevant by the model); and (2) statistical significance in Differential Gene Expression (DGE) analysis. This sequential two-gate design makes the SHAP-DGE-GSEA pipeline more conservative than standard DGE-GSEA: SHAP > 0 restricts the gene set to model-relevant genes, and DGE significance and GSEA enrichment significance ($p < 0.05$) jointly determine the biological conclusions. The SHAP threshold governs which genes enter the enrichment analysis as candidates; biological significance is established by GSEA NES significance, not by SHAP magnitude.

**TABLE 1** | Cross-study validation reveals the definitive hierarchy of mechanosensitive fascial cells.

| Cell type | Accuracy (%) | G-index | MR |
|---|---|---|---|
| CPTCs | 80.26 | +0.605 | 1 |
| Fibroblast | 71.77 | +0.435 | 2 |
| Macrophage | 53.15 | +0.063 | 3 |
| Lymphocyte | 42.11 | −0.158 | 4 |
| Endothelial | 30.50 | −0.390 | 5 |
| Mural cell | 26.97 | −0.461 | 6 |

*Note:* Cross-Study G-Index = (Accuracy−0.5)/0.5, measuring chance-corrected classification accuracy on an entirely independent fascia scRNA-seq dataset. Models were trained from scratch on the current study with hyperparameters optimized via GroupKFold inner cross-validation. MR, Mechanosensitivity Ranking. This table supersedes the within-study evaluation (Table S4) as the definitive ranking.

### 2.15 | Intercellular Communication Analysis

All intercellular communication analyses, including both intra-tissue and cross-tissue signaling inference, were performed using CellCall. For cross-tissue inference (Figure 6A), scRNA-seq data from fascial and colonic CPTCs were merged and analyzed jointly. The resulting interaction predictions represent computationally inferred potential signaling based on ligand–receptor co-expression patterns and should be interpreted as hypothesis-generating rather than as direct evidence of intercellular communication in vivo. PySCENIC analysis inferred transcription factor regulatory networks and activity scores.

### 2.16 | Cross-Tissue Cellular Homology Verification

To verify the homology between fascial and colonic CPTCs, the Seurat objects annotated as CPTCs from both tissues (encompassing both acupuncture and dysentery conditions) were extracted and merged using the merge function. The combined dataset underwent the unified preprocessing pipeline established in this study, including standard Seurat processing and batch effect correction via Harmony. Uniform Manifold Approximation and Projection (UMAP) was employed to visualize the transcriptomic alignment between the two cell populations. This process essentially identifies cross-batch "anchors" or biological correspondences. Distinct cell populations with fundamentally different transcriptomic profiles (e.g., colonic epithelial cells versus fascial CPTCs) would lack such matching anchors, thereby providing no biological basis for meaningful algorithmic alignment [9].

### 2.17 | Primary Isolation and Culture of Subcutaneous Fascia CPTCs

CPTCs were isolated from rat abdominal midline fascia following established protocols, aligning with methodologies outlined by Kota [10]. Tissue samples were dissected and digested using a collagenase type 4-based dissociation solution in PBS supplemented with 5% Fetal Bovine Serum (FBS). The resulting cell suspension was sequentially filtered through 100 μm and 40 μm sieves. Enzymatic activity was halted by washing with 10% FBS, and cells were initially plated onto T75 flasks for a brief 2-h adherence. Subsequently, slowly adhering cells were transferred to new flasks for continued culture.

The isolated CPTCs were cultured in DMEM/F12 medium, supplemented with 10% FBS and 1% penicillin–streptomycin, under standard conditions of 5% $CO_2$ at 37°C, maintaining 95% humidity. Immunofluorescence staining was employed to confirm cell purity. For further experimentation, cell slides were prepared in 12-well plates, with samples fixed using 4% paraformaldehyde at 4°C for 24 h after 72 h of culture.

### 2.18 | CCK-8 Cell Viability Assay

The CCK-8 assay was conducted to assess the effects of various concentrations of Salmonella on cell viability. Cell suspensions containing approximately 5000 cells in 100 μL per well were seeded into 96-well plates and incubated at 37°C with 5% $CO_2$ for 24 h. Following the incubation, the cells were treated with different concentrations of Salmonella for 20 h. After treatment, 10 μL of CCK-8 reagent was added to each well, and the plates were incubated for an additional hour. The absorbance at 450 nm was measured using a multifunctional enzyme reader, and the mean absorbance was calculated from three independent experiments.

### 2.19 | In Vitro Co-Culture Model and Bacterial Challenge Assay

To replicate an in vivo infection environment, NCM460 cells were assigned to three experimental groups: blank control (control), infection (Sal), and coculture-infection (CPTC-Sal). The NCM460 of each group was plated in triplicate into the lower chamber of transwell plates and cultured in DMEM/F12 medium supplemented with 10% fetal bovine serum (FBS). Upon reaching 80%–90% confluence, media for all groups was refreshed with FBS-supplemented DMEM/F12 medium, omitting penicillin and streptomycin. A Salmonella suspension at a multiplicity of infection (MOI) of 1 was introduced to the infection and coculture-infection groups.

In the co-culture setup, CPTCs were seeded in the transwell's upper chamber, allowing indirect interaction with NCM460 cells in the lower chamber. The cells underwent co-culture in antibiotic-free DMEM/F12 medium with 10% FBS for 20 h. Post-incubation, triplicate samples and cell lysates were harvested for subsequent analyses.

### 2.20 | Bacterial Colony Counting

To assess bacterial colonization, the NCM460 cell lysate was diluted in phosphate-buffered saline (PBS) to prepare a cell slurry. This slurry was then inoculated onto MacConkey agar plates supplemented with 50 μg/mL kanamycin and incubated at 37°C overnight. The colonies were subsequently counted to determine the colony-forming units (CFU). CFUs were compared between the infected and co-culture groups to evaluate the effect of telocyte co-culture on bacterial colonization.

### 2.21 | Statistical Analysis

Statistical comparisons employed two-tailed *t*-tests or Analysis of Variance (ANOVA). Multiple testing correction used the Benjamini-Hochberg method. $p$-values $< 0.05$ were considered statistically significant.

## 3 | Results

### 3.1 | Somatic Mechanotherapy Reverses Bacterial Colitis and Remodels the Fascia–Colon Single-Cell Landscape

To evaluate the therapeutic efficacy of somatic mechanotherapy at the macroscopic level, we first performed a comprehensive pathological and cellular assessment of rat colon tissues.

Hematoxylin and eosin (H&E) staining revealed severe pathological damage in the dysentery model group, characterized by disordered crypt architecture, significant submucosal edema, and massive infiltration of inflammatory cells in the lamina propria. In contrast, animals receiving acupuncture treatment exhibited marked improvement in colonic tissue structure, with restored crypt morphology, alleviated edema, and significantly reduced inflammatory infiltration, presenting a histological phenotype closer to the healthy state (Figure 1A).

We further assessed cell survival using the TUNEL assay. Colonic epithelial cells in the dysentery group showed extensive positive signals, indicating strong apoptosis induced by *Salmonella* infection, which directly compromised mucosal barrier integrity. Mechanotherapy significantly attenuated this apoptotic response. Quantitative analysis via CellProfiler demonstrated a highly significant reduction in the ratio of TUNEL-positive cells in the treated group compared to the dysentery group ($p < 0.001$), confirming its potent anti-apoptotic and barrier-protective effects (Figure 1B).

At single-cell resolution, we constructed a high-resolution single-cell atlas of the colonic tissue. Unsupervised dimensionality reduction via uniform manifold approximation and projection (UMAP) visualized a complex cellular ecosystem, categorizing cells into distinct clusters based on their transcriptomic profiles (Figure 1G). We annotated these populations using canonical lineage markers, including Cd34/Pdgfra, identifying CPTCs from other stromal subclusters or epithelial cells (Figure 1F). The remodeling of the colonic cellular landscape was evident. Cell proportion analysis showed a trend of expansion for epithelial populations and a concurrent reduction in immune subsets (lymphocytes and myeloid cells) in the acupuncture group (Figure 1C), consistent with the histological observation of reduced inflammation. Cell cycle analysis further revealed that mechanotherapy promoted epithelial proliferation, with a consistently elevated proportion of cells in the G2/M phase across multiple subsets, including enterocytes and goblet cells (Figure 1D). Moreover, pathological gene signatures associated with colitis (e.g., *Hp*, *Il1a*, *Ltf*) were broadly downregulated across multiple cell types (Figure 1E).

### 3.2 | Interpretable Deep Learning Identifies Fascial CPTCs as the Primary Mechanosensors via an AP-1/Hsp70 Axis

To unbiasedly identify the cell types most responsive to somatic mechanical stimulation, we applied the CARSS framework. This approach reframes "mechanosensitivity" as a machine learning "generalization performance" problem. We first constructed a comprehensive single-cell atlas of the subcutaneous fascia (Figure 2A) and clearly annotated the major cell types, including CPTCs, macrophages, and fibroblasts (Figure S8). To ensure the rigor of our single-cell analysis, we validated the in situ presence of CPTCs via immunofluorescence (Figure S2) and confirmed the absence of batch effects in our dataset (Figure S3). Initial visualization revealed a distinct transcriptomic shift in the CPTCs population between acupuncture and dysentery conditions, whereas other major populations (macrophages, fibroblasts) showed less separation (Figure 2A1), providing an initial cue for their responsiveness.

To establish a mechanosensitivity ranking free of training-data contamination, we performed cross-study validation using an entirely independent animal cohort (Methods 2.13.6). Table 1 reports the definitive G-Index values derived from this validation. Among six fascial cell types, CPTCs achieved the highest cross-study G-Index (+0.61; accuracy 80.3%), followed by Fibroblasts ($G = +0.44$; accuracy 71.8%) and Macrophages ($G = +0.06$; accuracy 53.2%). Endothelial cells, Mural cells, and Lymphocytes yielded G-Index $\leq 0$, indicating that their acupuncture-responsive transcriptional patterns failed to generalize beyond chance level to the independent cohort. SHAP analysis on the CPTC cross-study model independently identified Hspa1b as the top feature among all 19 910 genes (Table S3), confirming the biological consistency of the mechanosensitive signature across independent datasets. The cross-study validation provides an evaluation free of data leakage; Table 1 is therefore designated as the definitive mechanosensitivity ranking throughout this manuscript. The within-study evaluation (Table S4) provides complementary context but included training-partition data in its generalization assessment.

Within-study dual-evaluation (Self-matrix vs. Generalization testing) provided supporting evidence across 15 cell types (Table S4). Fascial CPTCs, macrophages, and fibroblasts emerged as the top-ranking responders, consistent with the cross-study validation. CPTCs achieved a training accuracy of 99.86% and a generalization accuracy of 86.32% in complex mixed datasets (Figure 2B). Fascia lymphocytes showed poor generalization (G-Index $> 1.0$), indicating weak specific responses in this within-study framework.

To decode the molecular features driving this classification, we employed SHAP (SHapley Additive exPlanations) analysis (Figure S1 describes the architecture). In fascial CPTCs, the top SHAP-ranked genes were highly enriched in "Response to mechanical stimulus" (GO:0009612, $NES = 1.70$, $p = 0.0008$) (Figure 2C,E and Table 2), while colonic CPTCs genes were enriched in tissue repair and chemokine functions, instead of any pathways directly responding to mechanical stimulus (Figure 2D,E). Specifically, the immediate-early genes [11] *Jun* and *Fos* (components of the AP-1 complex) and heat shock protein *Hspa1b* (Hsp70) were identified as core predictive features (Figure 2F). Differential expression and SCENIC analysis confirmed the upregulation of the AP-1 regulon in the acupuncture group (Figure 2F), hinting at an AP-1/Hsp70 axis as the primary mechanotransduction mechanism in fascia CPTCs. A comprehensive heatmap of the top 50 SHAP genes further illustrates the distinct molecular signatures driving the model's decision-making (Figure S5).

To further dissect the functional patterns of high-performing cell types, we employed an integrated SHAP-DGE-GSEA analysis method (Table 2). This approach filters differentially expressed genes based on their SHAP contribution scores, effectively removing noise introduced by linear computation methods through the non-linear learning capacity of DNN models [12]. The results revealed a critical functional differentiation among fascial cells following acupuncture intervention. Fascial CPTCs emerged as the sole cell population exhibiting significant positive enrichment in the "Response to mechanical stimulus" pathway (GO:0009612), while fascial macrophages and fibroblasts—despite being identified by CARSS as highly responsive to acupuncture—showed non-significant or negative enrichment for this pathway (Table 2). This indicates that macrophages and fibroblasts may not directly respond to mechanical stimulation but rather undergo "reprogramming" of their other functions by acupuncture.

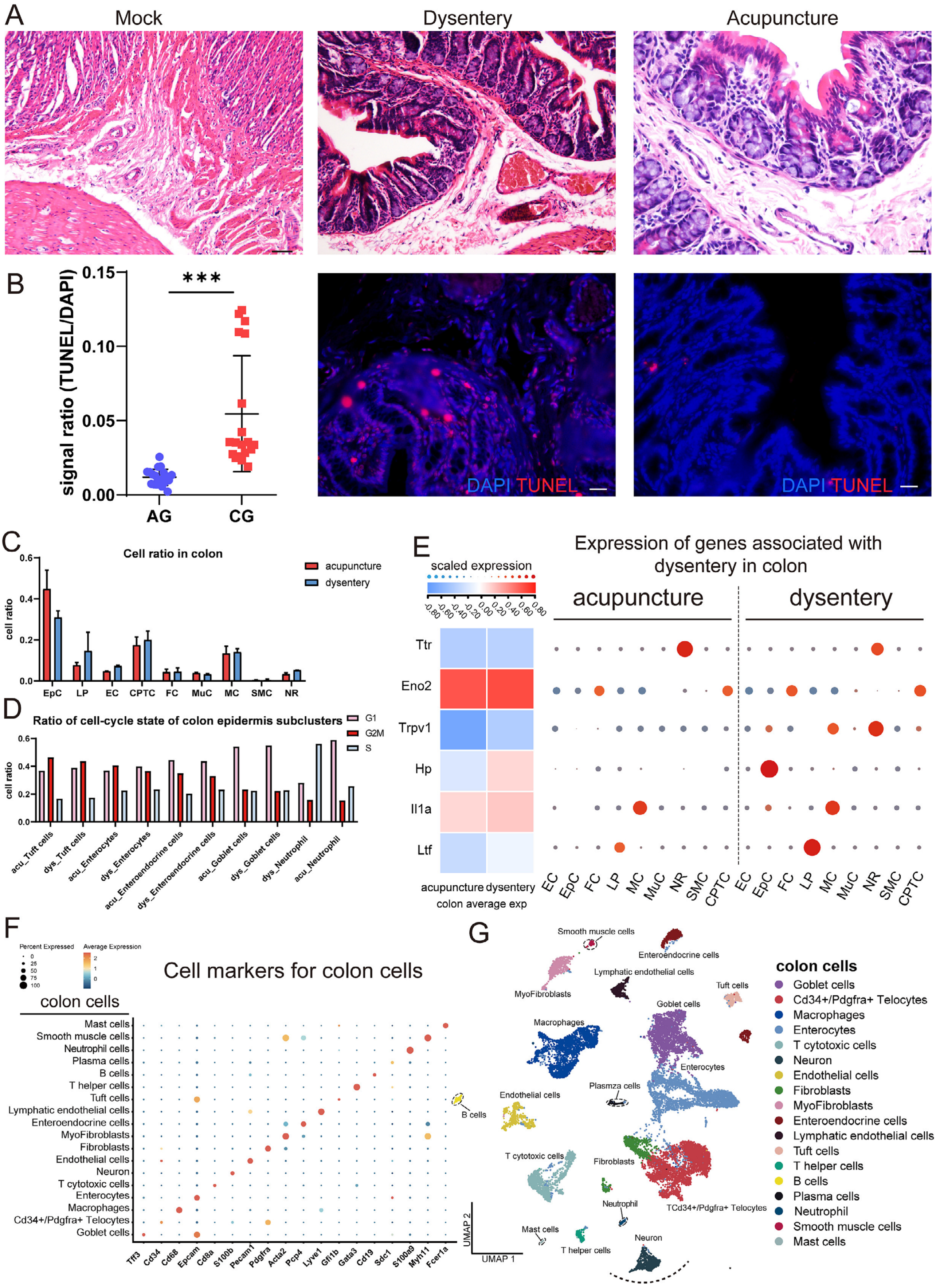

**FIGURE 1** | Legend on next page.

**FIGURE 1** | Somatic mechanotherapy reverses bacterial colitis and remodels the fascia–colon single-cell landscape. (A) Representative hematoxylin and eosin (H&E) stained sections of rat colon from Mock (healthy control), Dysentery, and Acupuncture-treated groups. The dysentery model group shows severe mucosal damage including crypt structural destruction and massive immune cell infiltration; acupuncture treatment significantly reverses these pathological changes, restoring tissue morphology closer to the healthy state. Scale bar: 40 μm. (B) Cell apoptosis assessed by TUNEL assay. Left: Quantification of TUNEL/DAPI signal ratio showing the acupuncture-treated group (AG) has significantly lower apoptosis rates than the dysentery control group (CG). Data presented as mean ± SEM. ***$p < 0.001$. Right: Representative immunofluorescence images show TUNEL-positive apoptotic cells (red) and DAPI-stained nuclei (blue). Scale bar: 20 μm. (C) Bar plot showing changes in major colon cell type proportions (Cell ratio in colon) based on scRNA-seq data. Compared to the dysentery group (blue), the acupuncture group (red) shows significantly increased epithelial cell (EpC) proportions and altered immune cell proportions, indicating reduced inflammation. CPTC, $CD34^+$ $PDGFR\alpha^+$ telocytes; EC, endothelial cells; EpC, epithelial cells; FC, fibroblasts; LP, lymphocytes; MC, macrophages; MuC, myofibroblasts; NR, neurons; SMC, smooth muscle cells. (D) Cell cycle state analysis of colon epithelial cell subpopulations (Ratio of cell-cycle state of colon epidermis subclusters). Bar plots show cell proportions in G1 phase (light color), G2/M phase (medium color), and S phase (dark color). Acupuncture treatment consistently increases the proportion of cells in proliferative phase (G2/M) across epithelial subtypes including Tuft cells, Enterocytes, Enteroendocrine cells, Goblet cells, and Neutrophils, suggesting enhanced tissue regeneration capacity. (E) Molecular signature of dysentery pathology (Expression of genes associated with dysentery in colon). Left heatmap shows scaled average expression levels of dysentery-associated genes (Ttr, Eno2, Trpv1, Hp, Il1a, Ltf) comparing acupuncture versus dysentery conditions, demonstrating overall downregulation after acupuncture. Right dot plot analyzes expression changes of these key genes at the single-cell type level across multiple cell populations, showing widespread reduction in expression in the acupuncture group. (F) Dot plot showing canonical marker gene expression used for colon cell type annotation (Cell markers for colon cells). Dot size represents the percentage of cells expressing the gene in each cluster, and color intensity corresponds to the scaled average expression level. Multiple cell types are identified including Goblet cells, $Cd34^+$/$Pdgfra^+$ Telocytes, Macrophages, Enterocytes, and others. (G) UMAP visualization of integrated single-cell transcriptomic data from all colon samples (acupuncture and dysentery groups), revealing a highly heterogeneous and diverse cellular ecosystem with distinct epithelial, immune, and stromal lineages clearly delineated. Cluster identities are annotated in the legend, including 19 distinct cell types.

Notably, within fascia CPTCs, canonical mechanosensory genes (Jun, Fos, Hspa1b) were significantly upregulated following acupuncture, whereas NF-κB family genes associated with inflammatory signaling were downregulated (Nfkb1: Log2FC = −0.43, $p = 8.97 \times 10^{-40}$; Table S1), arguing against a non-specific composite stress interpretation.

Further pathway enrichment comparison between DGE-GSEA and SHAP-DGE-GSEA methods (Table 3) revealed distinct molecular signatures for each cell type. Fascial fibroblasts showed significant changes in extracellular matrix remodeling pathways, suggesting acupuncture may regulate their fibrotic activity to maintain tissue homeostasis. Fascial macrophages exhibited significant negative enrichment in pro-inflammatory pathways (e.g., "Neutrophil migration", "Response to lipopolysaccharide", "Inflammatory response"), providing direct cellular-level evidence for the anti-inflammatory effects of acupuncture.

The cross-study G-Index values reported in Table 1 were derived from models trained from scratch on the current study and evaluated on an entirely independent fascia scRNA-seq dataset (independent cohort). SHAP analysis on the CPTC cross-study model independently identified Hspa1b as the top feature among all 19910 genes (Table S3), further confirming the biological consistency of the mechanosensitive signature across independent datasets.

### 3.3 | CPTCs Constitute a Central Communication Hub in the Colonic Mesenchymal Niche Under Homeostasis

Before analyzing the therapeutic effects, we established the baseline cellular landscape of the colon. Using a high-resolution annotation strategy, we defined 30 distinct cell subsets in the healthy colon (Figure 3A,C). Cell–cell communication analysis revealed that CPTCs function as a "busy" signaling hub in the healthy state, directing a vast number of signals to epithelial, macrophage, and lymphocytic populations (Figure 3B).

Hierarchical clustering of these communication patterns identified two primary signaling modules orchestrated by CPTCs: (1) an immunomodulatory module characterized by chemokine and inflammatory signals targeting immune cells (Figure 3D), and (2) a regenerative module dominated by Wnt and Fgf signals targeting epithelial stem/progenitor cells (Figure 3E). These findings confirm that under homeostatic conditions, CPTCs are critical regulators of both immune equilibrium and epithelial renewal.

### 3.4 | Mechanotherapy Reprograms the Colonic Immune Microenvironment and Transcription Factor Landscape

We next investigated how somatic stimulation alters this colonic network. Global communication analysis showed that acupuncture intensified the overall signaling flux, particularly enriching interactions between colonic CPTCs and macrophages (Figure 4A). SCENIC analysis revealed a shift in transcription factor activity: colonic CPTCs showed enhanced activity of the immunoregulatory factor *Irf7*, while *Nfkb1* activity was consistently downregulated across all immune cell types (Figure 4B,C), paralleling the resolution of inflammation observed histologically. To validate the functional consequences of this reprogramming, we confirmed that mechanotherapy actively restores tissue homeostasis by SHAP-GSEA analysis for colonic lymphocytes and epithelial cells (Figure S4).

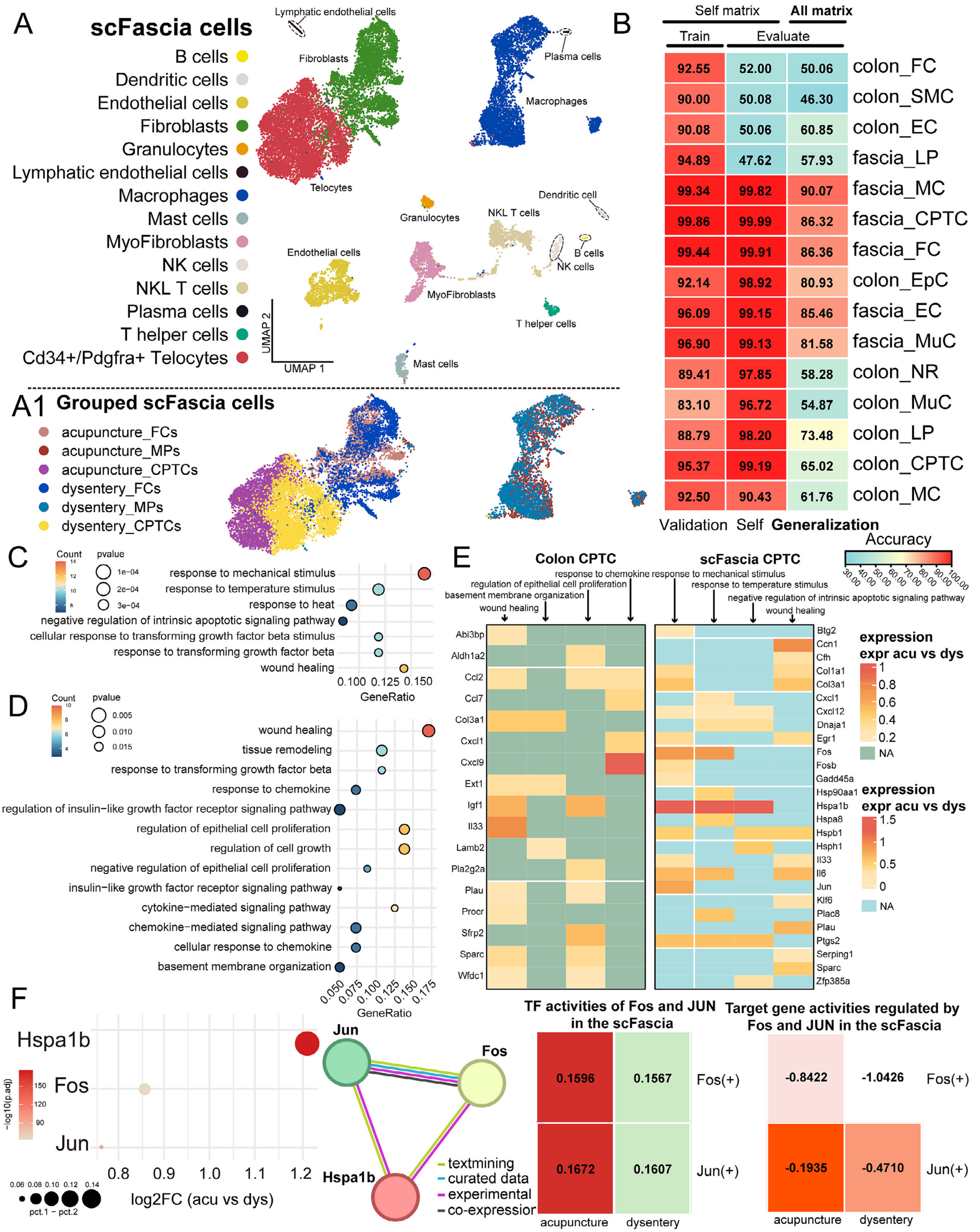

**FIGURE 2** | Legend on next page.

**FIGURE 2** | Interpretable deep learning identifies fascial CPTCs as the primary mechanosensors via an AP-1/Hsp70 axis. (A) UMAP visualization of the comprehensive subcutaneous fascia cellular atlas, identifying all major cell populations. (A1) UMAP plot highlighting the three most abundant cell types (CPTCs, macrophages, fibroblasts), grouped by treatment condition. A significant transcriptomic shift is observed in the acupuncture group's CPTC population, suggesting their high responsiveness to the treatment. (B) Performance evaluation of cell type-specific machine learning models (multilayer perceptrons) for distinguishing acupuncture from dysentery states. Models are evaluated on both accuracy within their own cell type ("Self-validation") and their ability to generalize across the entire tissue matrix ("Generalization"). The fascia CPTC model's excellent accuracy (99.99% self-validation, 86.32% generalization) powerfully identifies it as the most sensitive and reliable cell type for acupuncture. (C) Gene Ontology (GO) enrichment analysis of SHAP-ranked genes in fascial CPTCs reveals "Response to mechanical stimulus" as the most significantly enriched biological process, providing direct computational evidence for their role as primary mechanosensors. (D) In distal colonic CPTCs, SHAP-ranked genes are primarily associated with tissue homeostasis pathways such as "wound healing" and "tissue remodeling," demonstrating the systemic therapeutic consequences of initial mechanical sensing. (E) Comparative heatmap showing expression of top SHAP-ranked genes in colonic CPTCs (left) and fascial CPTCs (right). This illustrates that gene programs activated by acupuncture at local (mechanical sensing) and distal (tissue repair) sites are distinct yet functionally related. (F) Mechanistic dissection of the AP-1 stress response axis in fascial CPTCs. Differential expression analysis confirms significant upregulation of Hspa1b (Hsp70), Jun, and Fos in acupuncture response. Protein–protein interaction network visualizes their known functional connections. Critically, PySCENIC analysis confirms enhanced transcriptional activity of the AP-1 complex and its target gene program in the acupuncture group, establishing the AP-1-Hsp70 axis as the core mechanotransduction pathway in CPTCs.

**TABLE 2** | SHAP-DGE-GSEA analysis reveals hierarchical mechanical response in fascial cells.

| Cell type | Analysis method | NES | $p$ | Significant |
|---|---|---|---|---|
| Fascial CPTCs | DGE-GSEA | 1.65 | 0.000567 | Yes |
| | **SHAP-DGE-GSEA** | **1.70** | **0.000839** | **Yes** |
| Fascial Mural cells | DGE-GSEA | — | — | — |
| | **SHAP-DGE-GSEA** | **1.31** | **0.056** | **No** |
| Fascial Endothelial cells | DGE-GSEA | 1.12 | 0.219 | No |
| | **SHAP-DGE-GSEA** | **1.11** | **0.247** | **No** |
| Fascial Fibroblasts | DGE-GSEA | −1.36 | 0.280 | No |
| | **SHAP-DGE-GSEA** | **−1.37** | **0.281** | **No** |
| Fascial Macrophages | DGE-GSEA | −1.31 | 0.312 | No |
| | **SHAP-DGE-GSEA** | **−1.30** | **0.329** | **No** |

*Note:* Bold formatting indicates SHAP-DGE-GSEA method results. This table shows the hierarchical relationship of different fascial cell types in response to mechanical stimulus (GO:0009612). CPTCs exhibit the strongest positive enrichment, while macrophages and fibroblasts show negative enrichment without statistical significance. SHAP filtering enhances the response signal in CPTCs, further highlighting their role as the primary mechanosensors.
Abbreviation: NES, Normalized Enrichment Score.

### 3.5 | Resolution of Inflammatory CPTC Subsets and Emergence of a Wnt-Driven Regenerative Network

To dissect the specific signals driving this repair, we subclustered the colonic cells from acupuncture/dysentery groups into 33 functional subsets (Figure 5A,F and Figure S6 acts as the annotation dictionary). First, we counted the number of sent and received signals in the overall cell communication pattern of the colon, and the results showed that, compared to the number of signals sent, the number of signals received by colon cells showed greater fluctuation under acupuncture intervention, especially for colon CPTC, macrophages, and proliferative epithelial cells (Figure 5B). Specifically, acupuncture intervention significantly (a difference of more than 30% signals received) increased the number of received signals for inflammatory CPTCs and matrix secretory CPTCs, while decreasing the number of received signals for tissue-repair CPTC; at the same time, acupuncture dramatically reduced the number of received signals for antigen-presenting, inflammatory, and simple inflammatory-resistant macrophages, while increasing the number of received signals for macrophages that combine anti-inflammatory and tissue-repair functions, and the number of signals sent by this type of macrophage also showed a slight increase after acupuncture (Figure 5D). In the cell communication analysis framework of this study, receiving a signal indicates the regulatory effect of intracellular transcription factors (TFs) and downstream regulator gene expression; therefore, the above results strongly suggest that acupuncture intervention influences intestinal cellular functions—namely, enhancing colonic inflammatory regulation and tissue repair capacity.

Next, we observed a dramatic remodeling of the CPTC-Macrophage interactome. In the dysentery state, a specific Ptgs2$^{+}$ (COX-2) inflammatory CPTC subset was prevalent, driving a pathological loop by secreting Il-6 to maintain M1-like inflammatory macrophages (Figure 5E and Figure S6), and mechanotherapy abolished this Il-6 signaling to M1-like inflammatory macrophages (Figure 5E). Strikingly, it promoted the emergence of regenerative CPTC subsets that upregulated Wnt agonist

**TABLE 3** | Comparison of pathway enrichment analysis between DGE-GSEA and SHAP-DGE-GSEA methods (top 3 most significant pathways).

| Cell type | Pathway ID | Pathway name | DGE-GSEA NES (*p*) | SHAP-DGE-GSEA NES (*p*) | ΔNES |
|---|---|---|---|---|---|
| Fascial CPTC | GO:0009612 | Response to mechanical stimulus | 1.65 (0.000567) | 1.70 (0.000839) | +0.05 |
| | GO:0009408 | Response to heat | 2.23 (0.00014) | 2.27 (0.00017) | +0.04 |
| | GO:0006458 | *De novo* protein folding | N/A | 1.86 (0.031) | New |
| Fascial fibroblast | GO:0030198 | Extracellular matrix organization | −1.28 (0.201) | −1.67 (0.045) | −0.39 |
| | GO:0030199 | Collagen fibril organization | 1.21 (0.25) | 1.84 (0.031) | +0.63 |
| | GO:0006119 | Oxidative phosphorylation | 2.11 (0.00001) | 2.11 (0.00001) | 0 |
| Fascial macrophage | GO:1990266 | Neutrophil migration | −2.29 (0.001) | −2.41 (0.001) | −0.12 |
| | GO:0032496 | Response to lipopolysaccharide | −2.24 (0.001) | −2.25 (0.001) | −0.01 |
| | GO:0006954 | Inflammatory response | −2.01 (0.001) | −2.02 (0.001) | −0.01 |
| Fascial endothelial cell | GO:0001525 | Angiogenesis | 1.78 (0.032) | 1.82 (0.029) | +0.04 |
| | GO:0007155 | Cell adhesion | −1.45 (0.089) | −1.52 (0.067) | −0.07 |
| | GO:0006119 | Oxidative phosphorylation | 2.34 (0.00001) | 2.35 (0.00001) | +0.01 |
| Fascial mural cell | GO:0030198 | Extracellular matrix organization | −1.42 (0.156) | −1.58 (0.089) | −0.16 |
| | GO:0006119 | Oxidative phosphorylation | 2.18 (0.00001) | 2.19 (0.00001) | +0.01 |
| | GO:0007155 | Cell adhesion | −1.23 (0.189) | −1.35 (0.145) | −0.12 |

*Note:* Positive values indicate pathway activation (positive enrichment), negative values indicate pathway suppression (negative enrichment). $p$-value $< 0.05$ is considered statistically significant. "DGE-GSEA" refers to classic GSEA based on differentially expressed genes; "SHAP-DGE-GSEA" refers to GSEA method incorporating SHAP feature importance weighting. ΔNES indicates the change in NES between the two methods.
Abbreviation: NES, Normalized Enrichment Score.

signaling (Figure 5C). Ligand-receptor analysis showed a shift from inflammatory chemokines (Ccl, Cxcl) activating Nfkb1 in macrophages to signals activating Rela, a subunit associated with cell survival and tissue repair (Figure 5G,H). Simultaneously, CPTCs increased the secretion of canonical (Wnt2) and non-canonical (Wnt5a, Wnt11) Wnt ligands targeting epithelial cells (Figure 5C), directly supporting the regenerative phenotype.

### 3.6 | A Systemic Pro-Repair Wnt Milieu Accompanies Fascial and Colonic CPTC Transcriptional Synchronization

A key question remains: how does physical force at the body surface transmit to the deep colon? We fused the scRNA-seq data of fascial and colonic CPTCs and observed a striking "transcriptional resonance"—despite their anatomical distance, they shared a highly conserved transcriptional trajectory and Wnt signaling profile (Figure 6D, see also Figure S7A,B). This systemic synchronization was quantitatively supported by a positive correlation ($R = 0.75$) in the transcriptional fold-changes of key mechanosensitive genes between the two tissue sites (Figure S7D). Notably, the mechanical alarmin Il33 and the regenerative ligand Wnt2 exhibited a concordant upregulation pattern in both fascial and colonic CPTCs (Figure S7C), suggesting that a conserved transcriptional program may operate across these anatomically distant CPTC populations.

We hypothesized a systemic bridge. CellCall analysis of the "Fascia-to-Colon" axis predicted a robust Wnt signaling flow from fascial CPTCs to colonic CPTCs, mediated by *Wnt2*/*Wnt10b* ligands (Figure 6A). In parallel, transmission electron microscopy (TEM) of the acupoint fascia revealed active secretion of high-electron-density extracellular vesicles (EVs) from CPTCs into the microvascular vicinity (Figure 6B). Furthermore, ELISA confirmed a significant elevation of Wnt2 protein in the serum of acupuncture-treated rats compared to controls ($p < 0.001$) (Figure 6C). Viewed together, these observations—predicted Wnt connectivity between fascial and colonic CPTCs, EV-associated ultrastructure at the perivascular fascia, and elevated circulating Wnt2 (Figure 6A–C)—collectively support a model in which acupuncture establishes a systemic, pro-repair Wnt milieu that may coordinate transcriptional reprogramming across anatomically distant CPTC populations. While it is worth noting that we don't need to assume that every Wnt molecule that reaches the colon must come from the fascia, almost all metazoan cells secrete Wnt signal; serum is actually a confluence pool for this signal [13], and elevated serum Wnt2 indicates that the body has entered a pro-regenerative state due to mechanical therapeutic intervention.

Furthermore, this systemic synchronization was not limited to Wnt ligands but encompassed a broad spectrum of tissue-reparative modules (Figure S7). Based on scRNA-seq data from fascial and colonic CPTCs only, we identified a concordant upregulation of the mechanical alarmin Il33 [14] and the epithelial mitogen Fgf7 [15] in both CPTC populations, consistent with a conserved transcriptional program across the two sites. Conversely, the pro-apoptotic kinase Dapk1 [16] was synchronously downregulated in both sites (Log2FC $< 0$), providing a molecular explanation for the reduced epithelial apoptosis observed in Figure 1B. Across these seven functionally selected genes, a positive correlation in fold-change magnitude was observed between fascial and colonic CPTCs ($R = 0.75$, $p = 0.052$), supporting a directional concordance in their transcriptional response to acupuncture—suggestive of a conserved cross-tissue transcriptional program.

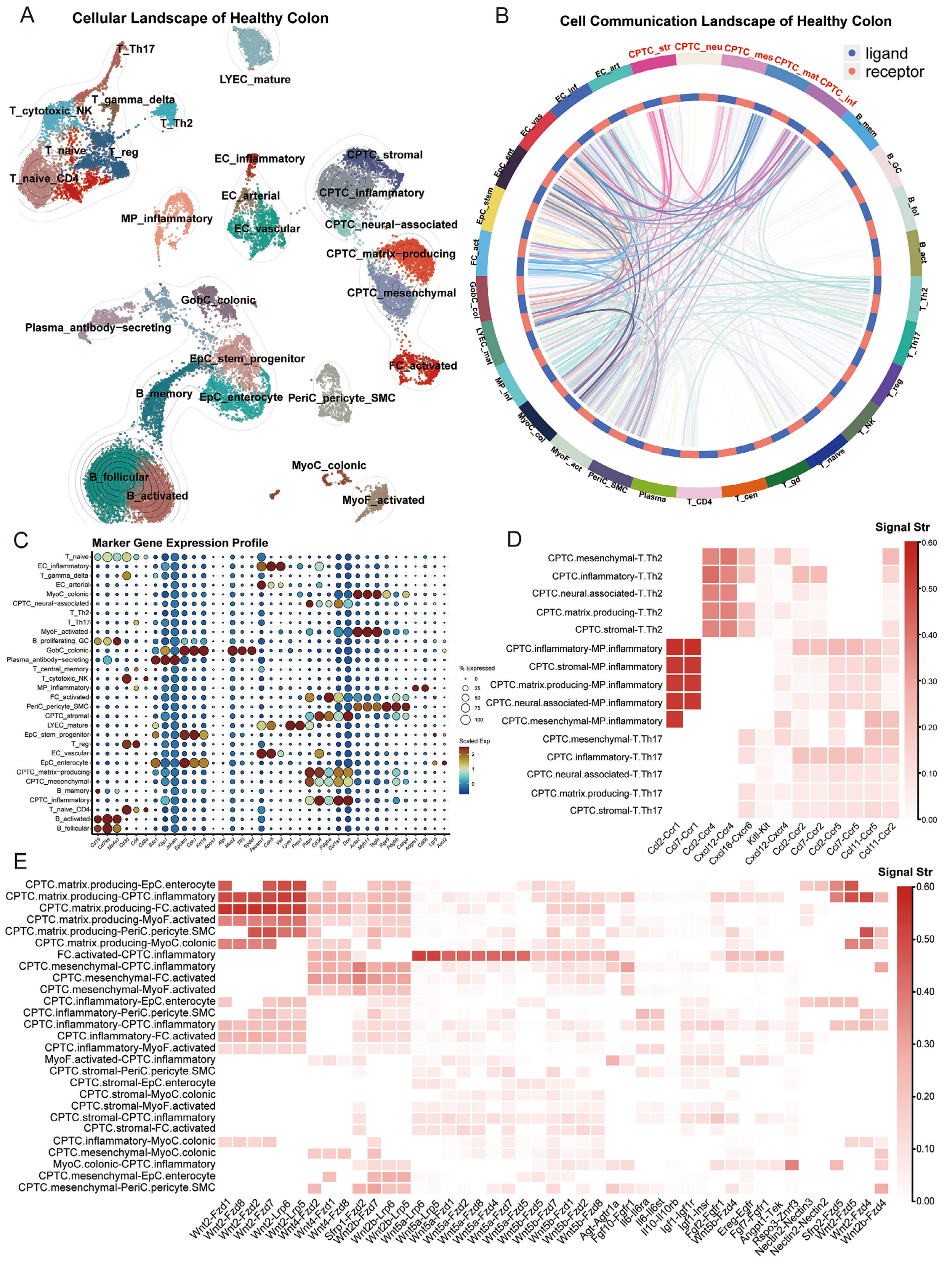

**FIGURE 3** | Legend on next page.

**FIGURE 3** | CPTCs constitute a central communication hub in the colonic mesenchymal niche under homeostasis. (A) UMAP visualization of scRNA-seq data from healthy mouse colon tissue (GSE264408). Analysis identifies 30 distinct cell clusters, including multiple colonic CPTC subpopulations along with other stromal, immune, and epithelial cell types. (B) Chord diagram displaying the global cell communication landscape. CPTC subpopulations (red text) serve as developed signaling hubs, sending large amounts of signals to other cell populations (such as epithelial cells, macrophages, lymphocytes, and fibroblasts). Blue and red arcs represent ligand and receptor interactions respectively, with line depth indicating signal strength—darker colors indicate stronger signals. (C) Marker gene expression profiles used for high-resolution annotation. Bubble size indicates the percentage of cells expressing the gene, and color scale represents scaled average expression level. (D) Heatmap showing ligand-receptor interactions between CPTC subpopulations and immune cells (such as T cells, macrophages), containing some cluster-specific signals. Communication patterns are characterized by chemotactic and inflammatory signals, particularly involving the Ccl and Cxcl ligand families. (E) Heatmap depicting signal interactions between CPTC subpopulations and colonic epithelial cells. Enrichment of Wnt and Fgf signaling pathways (such as Wnt2-Fzd, Fgf7-Fgfr) suggests CPTC's potential regulatory role in epithelial proliferation, differentiation, and stem cell maintenance.

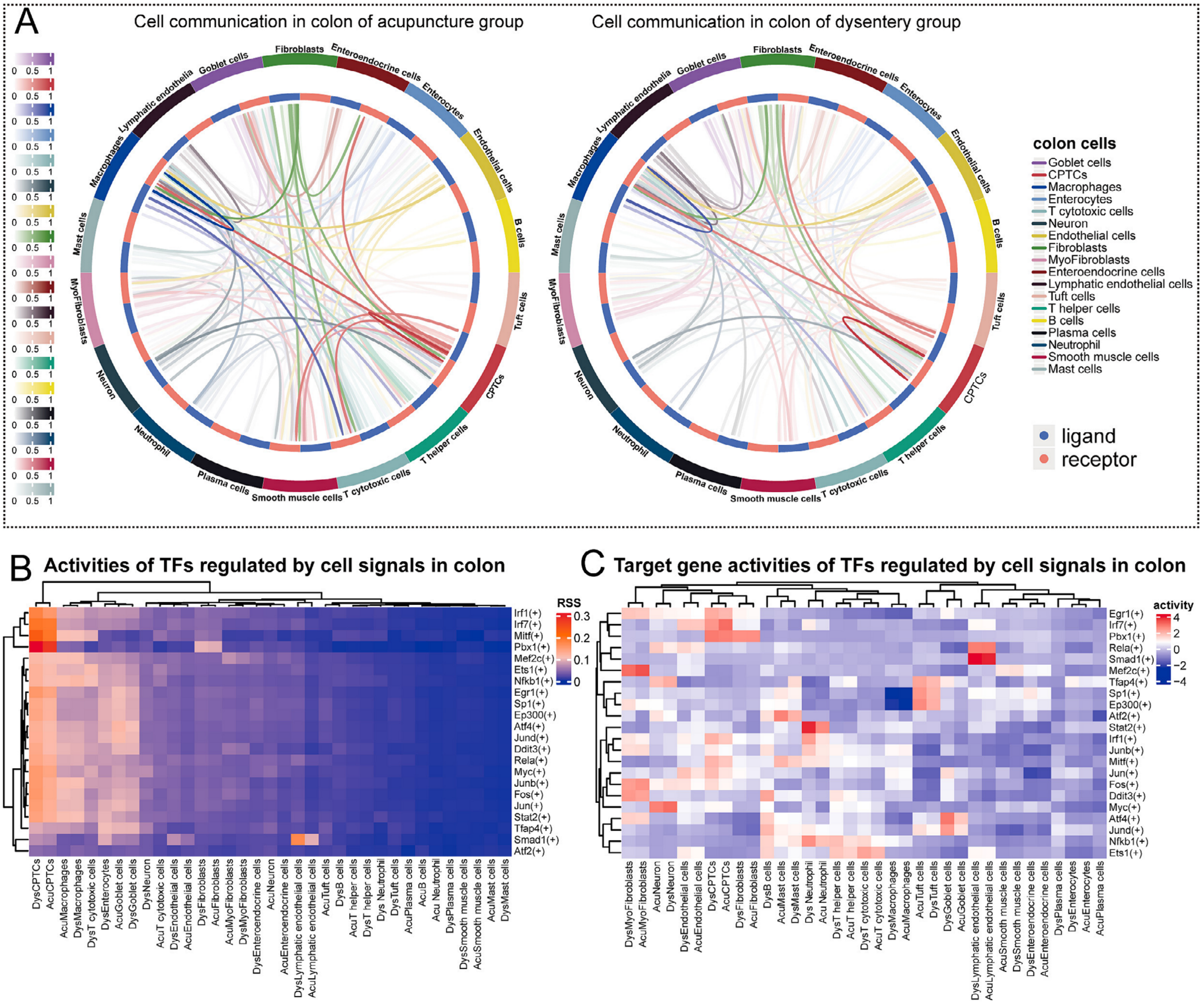

**FIGURE 4** | Mechanotherapy reprograms the colonic immune microenvironment and transcription factor landscape. (A) Chord diagrams of cell communication networks within the colon for acupuncture (left) and dysentery (right) groups. Acupuncture intervention enhances the overall intensity and complexity of intercellular signaling, particularly strengthening communication between CPTCs and immune cells (especially macrophages). (B) PySCENIC-inferred TF regulon RSS heatmap. Higher RSS values indicate highly cell type-specific TF regulation. CPTCs and macrophages exhibit unique transcriptional regulatory activities, characterized by high specificity for immune-related TFs (such as Irf7, Nfkb1) and repair-related factors (such as Myc). (C) PySCENIC-inferred TF regulon (target gene set) relative transcriptional activity heatmap. Color scale represents activity level (red: High; blue: Low). Acupuncture significantly increases Irf7 regulon activity in CPTCs, indicating enhanced immunomodulatory capacity; Nfkb1 regulon activity shows a consistent declining trend across all immune cell types, presumably related to inflammation reduction. Additionally, Stat2 regulon differences manifest as downregulation in acupuncture group neutrophils and activity upregulation in macrophages.

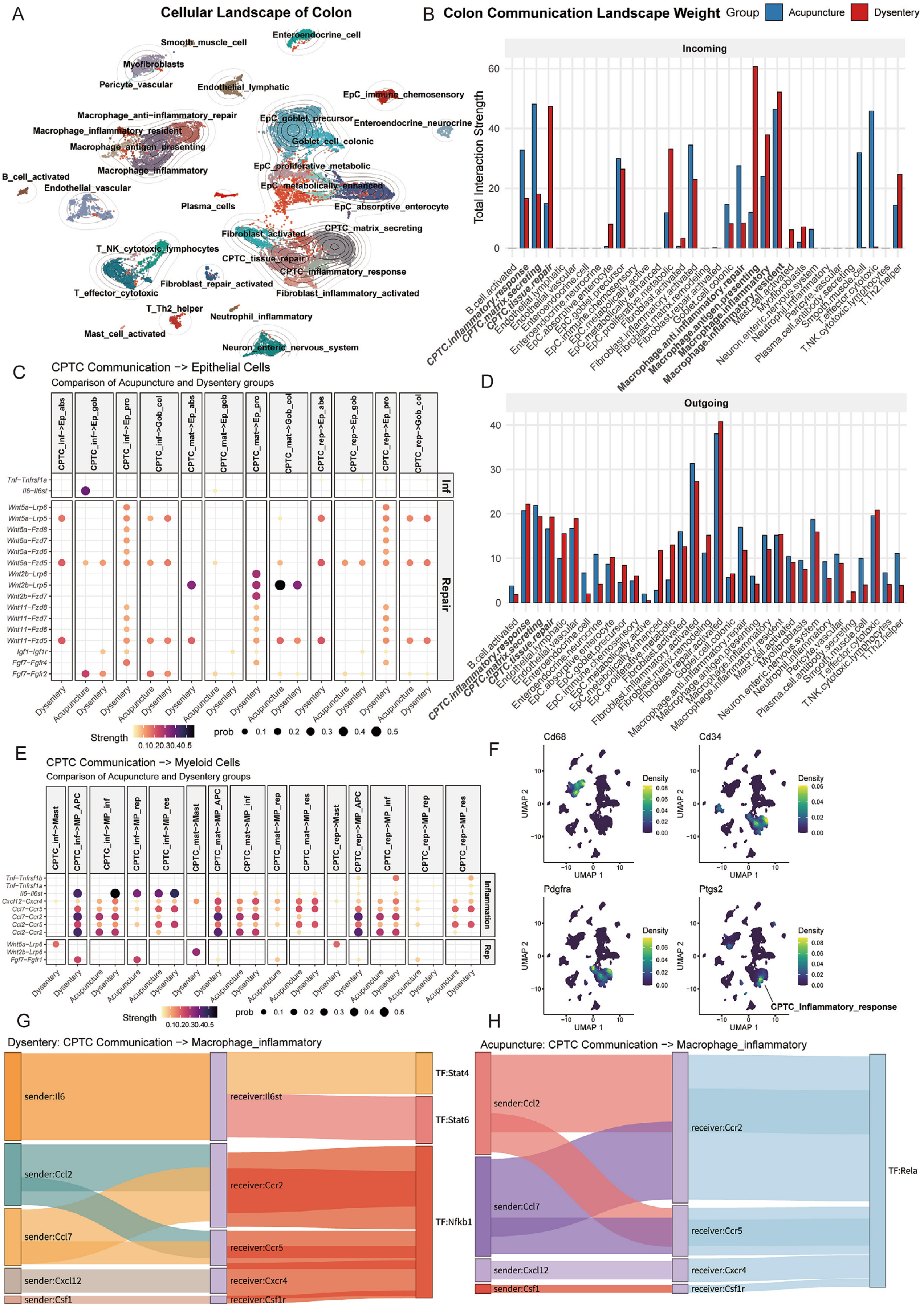

**FIGURE 5** | Legend on next page.

**FIGURE 5** | Resolution of inflammatory CPTC subsets and emergence of a Wnt-driven regenerative network. (A) High-resolution annotated UMAP of 33 cell subpopulations within acupuncture and dysentery group colons. (B, D) Bar plots of incoming and outgoing signal counts for each cell type. Compared to the dysentery group (red bars), acupuncture intervention (blue bars) alters the communication landscape. (C) Bubble plot of ligand-receptor interactions between CPTC subpopulations (senders) and epithelial cells (receivers). Enrichment of repair-related signaling pathways between the two cell types includes canonical (Wnt2) and non-canonical (Wnt5a, Wnt11) Wnt signals, as well as Fgf signals. (E) Bubble plot comparing ligand-receptor interactions between CPTC subpopulations and myeloid cells. Communication between these cell types is mainly dominated by inflammatory regulatory factors (Ccl, Cxcl, Il, Tnf). Il6 signaling from inflammatory-responsive CPTCs targeting inflammatory macrophages is significantly present in the dysentery group but disappears after acupuncture intervention. (F) Gene density plots showing key marker expression: Cd68 (macrophages), Cd34 and Pdgfra (CPTCs/stromal cells), and Ptgs2 (inflammatory-responsive CPTCs). (G, H) CellCall Sankey diagrams containing complete ligand-receptor-TF chains. In the dysentery group, Il6 signals from CPTCs activate Stat4 and Stat6 in inflammatory macrophages, while chemokine signals (Ccl, Cxcl) activate Nfkb1. In the acupuncture group, Il6-mediated pathways are eliminated, and chemokine signals mainly activate transcription factor Rela.

### 3.7 | Fascial CPTC Secretome Directly Rescues Epithelial Barrier Integrity via Wnt/β-Catenin/Myc Signaling

Finally, to verify causality and confirm that fascial CPTC-derived signals are sufficient to drive repair, we established a transwell co-culture system. Primary fascial CPTCs were seeded in the upper chamber, and colonic epithelial cells (NCM460) challenged with *Salmonella* were placed in the lower chamber, sharing only the fluid environment (Figure 7A,J).

The inclusion of fascial CPTCs significantly inhibited bacterial colonization in epithelial cells (Figure 7C) and dramatically reduced infection-induced apoptosis (TUNEL assay, $p < 0.0001$) (Figure 7D–F). At the molecular level, the CPTC secretome suppressed *TLR4*, TNF-α, and *IL6* expression while robustly upregulating the Wnt target genes *CTNNB1* (β-catenin) and *MYC* in the epithelial cells (Figure 7G–I). Transcription factor enrichment analysis (TFHomer) confirmed that the upregulated genes were enriched for c-Myc targets, which drove cell cycle progression and proliferation (Figure 7L,M). In addition, SCENIC analysis for scRNA-seq of intestinal epithelial cells showed changes in transcription factor regulatory activity of Egr2(+), Myc(+), and Sp1(+) before and after acupuncture intervention. The downstream genes of these transcription factors showed an overall upward trend in transcriptional activity in the epithelium and goblet cells after Mechanotherapy (Figure 7N). These results demonstrate that the CPTC secretome is sufficient to activate epithelial regeneration and restore barrier integrity through paracrine Wnt/β-catenin/Myc signaling.

## 4 | Discussion

In traditional Chinese medicine research, although abundant classical texts and empirical data are available for reference, there is often a lack of prior knowledge grounded in modern biological theory. We frequently lack prior knowledge linking signals to functions, and identifying specific cellular executors remains challenging. The CARSS framework proposed in this study transforms the biological identification problem into a machine learning "generalization performance evaluation" problem, successfully identifying key cell subpopulations responsive to acupuncture intervention under unsupervised conditions. This differs from both traditional hypothesis-driven biological research paradigms and conventional linear data analysis methods based on differential expression analysis. Compared to traditional differential gene expression (DGE) analysis, the deep learning approach combined with SHAP-based gene feature interpretation can capture non-linear interactions between genes and correct biological noise bias introduced by linear analysis algorithms [17], thereby more precisely screening authentic cellular response signatures. This powerfully corroborates the view that "Telocytes are potential initial mechanostimulus-responsive cells" [5]. This "data-driven" research paradigm provides an effective high-resolution, unbiased computational strategy for deciphering the biological mechanisms of complex interventions such as acupuncture.

Traditional perspectives on the biological effects triggered by acupuncture mechanical stimulation have typically focused on conventional fibroblasts or nerve endings in acupoint regions [18, 19]. However, computational biology analysis based on the CARSS framework's Generalization Index (G-Index) revealed a previously unnoticed situation: following acupuncture intervention, the transcriptomic signatures of fascial CPTCs underwent the most specific drift. DNN models trained on CPTC expression matrices achieved recognition accuracy in the same top tier as fibroblasts and macrophages. Most importantly, among all cell types in this top tier, only CPTCs showed significant positive enrichment in mechanosensory response functions based on expression profile changes, while other cells predominantly exhibited indirect biological functions following acupuncture intervention. Since the subcutaneous fascial CPTCs identified in this study are highly similar to the subcutaneous fascial Telocytes mentioned in previous research, this computational biology evidence strongly supports the hypothesis that CPTCs serve as "first responders" at acupoints [5].

While G-Index alone ranks transcriptional specificity without resolving its upstream cause, the distinction between direct mechanosensing and secondary effects was addressed through SHAP-DGE-GSEA functional annotation (Table 2).

Among all top-ranked cells, only CPTCs exhibited significant positive enrichment in the "Response to mechanical stimulus" pathway (GO:0009612, NES = 1.70, $p = 0.000839$), while macrophages and fibroblasts—despite high G-Index scores—showed non-significant negative enrichment, indicating their responses reflect secondary functional remodeling rather than primary mechanosensing. The AP-1/Hsp70 molecular signature identified by SHAP analysis is mechanistically consistent with known immediate-early responses to physical force, and is distinct from transcriptional programs associated with inflammation resolution or paracrine reprogramming. Definitive exclusion of secondary contributions awaits isolated mechanical loading controls, which represent a priority for future work. Notably, the NF-κB downregulation reported here is derived from a controlled comparison between acupuncture and dysentery groups—both exhibiting active inflammation—and therefore reflects the net effect of needle stimulation rather than spontaneous resolution of inflammation. We acknowledge that NF-κB suppression is not exclusive to mechanical loading—it also occurs during inflammation resolution and in response to cytokine signals that are present in this experimental model regardless of needle stimulation. The Mechanospecificity Index, while descriptively informative, is derived from gene sets and data within the same experiment and thus does not constitute an independent test of the mechanical versus inflammatory hypothesis. The definitive establishment of CPTC mechanosensory specificity will require isolated mechanical loading controls (e.g., ex vivo fascial stretch assays in the absence of inflammatory context), which we identify as a priority for future investigation.

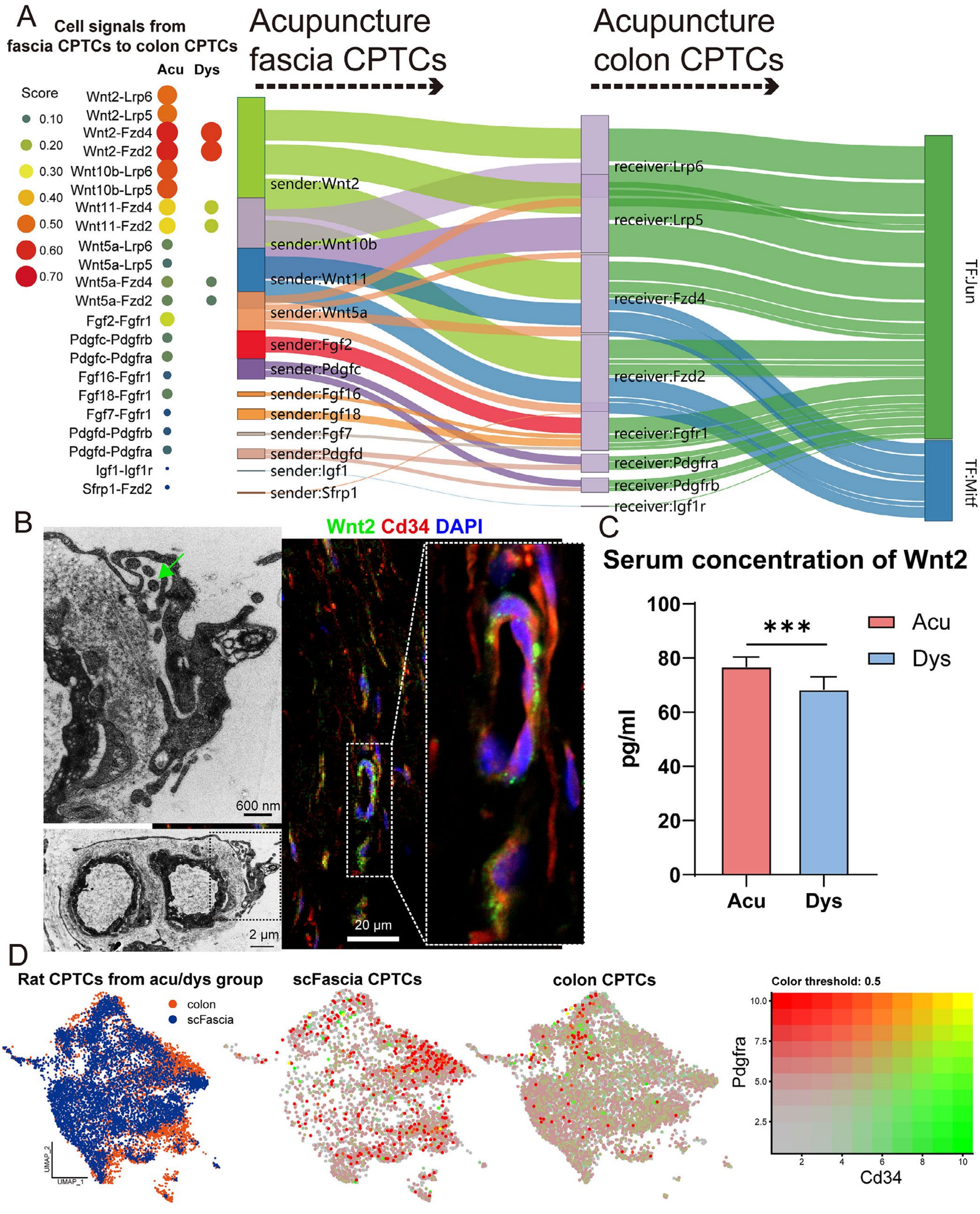

**FIGURE 6** | Legend on next page.

**FIGURE 6** | A systemic pro-repair Wnt milieu accompanies fascial and colonic CPTC transcriptional synchronization. (A) Cell communication inference analysis (CellCall). Left bubble plot shows interaction strength of signals sent from subcutaneous fascia CPTCs (senders) to colonic CPTCs (receivers) under acupuncture (Acu) and dysentery (Dys) conditions, with significantly enhanced Wnt signaling pathway (such as Wnt2-Fzd2) in the acupuncture group. Right Sankey diagram visualizes specific signal flow in the acupuncture group: Wnt ligands (Wnt2, Wnt10b, etc.) from fascial CPTCs primarily target receptors (Fzd2, Lrp5, etc.) on colonic CPTCs and are predicted to activate downstream transcription factors Jun and Mitf. Predicted signaling flows are computed inferences based on ligand–receptor expression profiles and do not constitute direct measurement of intercellular communication. (B) Ultrastructure and molecular features of the subcutaneous fascia microenvironment. Left: Transmission electron microscopy (TEM) shows perivascular CPTCs secreting high-electron-density extracellular vesicles (green arrows), scale bar: 600 nm (top)/2 μm (bottom). Right: Immunofluorescence staining shows spatial co-localization of Wnt2 protein (green) and CPTC marker Cd34 (red) around subcutaneous fascia microvasculature, with nuclei counterstained by DAPI (blue). (C) ELISA results show acupuncture-induced elevation of systemic Wnt signal abundance, with significantly elevated serum Wnt2 protein concentration in acupuncture-treated (Acu) rats compared to dysentery model (Dys) group (****$p < 0.001$*). (D) Cross-tissue single-cell data fusion analysis. Left: UMAP projection shows subcutaneous fascia (blue) and colon (orange) derived CPTCs exhibiting high topological overlap in transcriptional space. Middle/Right and legend: Feature plots display Pdgfra and Cd34 co-expression patterns in CPTCs from both tissue sources.

The convergence of CARSS-based identification with independent SHAP interpretation (AP-1/Hsp70 axis) and functional enrichment analysis (GO:0009612) across two independent analytical pipelines provides cross-method validation for the robustness of the CPTC mechanosensitive signature. Formal external validation on independent scRNA-seq datasets from other animal models or acupuncture paradigms remains an important future direction.

A methodological consideration regarding the cross-study validation is the asymmetric performance observed between transfer directions. In the reverse direction (current study → independent cohort), CPTCs achieved $G = +0.61$, Fibroblasts $G = +0.44$, and Macrophages $G = +0.06$, while Endothelial cells, Lymphocytes, and Mural cells showed $G \leq 0$; in the forward direction (independent cohort → current study), all cell types yielded $G \approx 0$. This asymmetry is consistent with directionally non-uniform batch effects between the two independently conducted experiments. Importantly, the same batch effects that suppress forward transfer can only attenuate—not inflate—reverse-direction performance. Therefore, the $G = +0.61$ observed for CPTCs in the reverse direction represents a conservative estimate of their true cross-batch signal strength, supporting an "informative rather than concerning" interpretation of this asymmetry (full forward and reverse data in Table S2). Model interpretation based on SHAP value ranking of genes identified *Jun* and *Fos* as the core signature genes driving CPTC state transitions. This echoes a previous biomechanical study on human periodontal ligament fibroblasts (hPDLFs): when primary hPDLFs were extracted for adherent culture, simple mechanical stretching alone could activate the MAPK signaling pathway within 15 min, while the transcription factor genes *Jun* and *Fos*, which encode the AP-1 protein complex, simultaneously reached peak expression at 1 h post-stretch and returned to baseline by 3 h post-stretch [20]. This study characterized the AP-1 transcription factor complex composed of Fos and Jun as an "immediate-early response" signature of cells responding to mechanical strain. Mechanical force has been demonstrated to directly induce rapid AP-1 phosphorylation and nuclear translocation, subsequently initiating downstream gene transcription [11, 20]. In our study, the enhanced AP-1 function exhibited by CPTCs suggests that the physical stress generated by acupuncture may be converted into intracellular transcriptional programs through similar mechanochemical transduction mechanisms, thereby completing the "primary encoding" of acupuncture signals.

It should be noted that due to technical limitations in early studies, the classic periodontal ligament fibroblast (PDLF) population mentioned above was a concept that long failed to distinguish between Telocytes and fibroblasts. The two are difficult to differentiate under light microscopy during in vitro culture, and marker systems were not yet clearly defined. Telocytes, as a cell population with outstanding mechanosensory and cell communication capabilities, have distinct single-cell expression profiles from periodontal ligament fibroblasts and have been identified as an independent cell population based on $Cd34^{+}/Pdgfra^{+}$ and other factor expression [21].

A core challenge facing this study was explaining how acupoint stimulation at the body surface alleviates inflammation in distant organs. In the dysentery model, the colonic cell communication pattern presented a pathological communication loop originating from Ptgs2+ CPTCs that maintains M1-like macrophages in a pro-inflammatory state through IL-6 signaling. This finding both echoes our conclusion from body surface acupuncture research that "Ptgs2+ CPTCs are the primary responders to acupuncture" [6] and aligns with a recent study on the "inflammatory imprint" of stromal cells in inflammatory microenvironments [22]. Correspondingly, acupuncture intervention clearly blocked pro-inflammatory signaling within the colon and induced colonic CPTCs to transition toward a tissue repair phenotype, manifested as enhanced activity of immunomodulatory factors such as Irf7 and downregulation of Nfkb1 activity in macrophages. This indicates that acupuncture does not simply suppress immune cell function, but rather actively alters the immune tone of the intestinal microenvironment by reshaping the signaling secretome of colonic stromal cells (such as CPTCs), promoting M1-like macrophages to reduce their inflammatory phenotype while enhancing the activity of macrophages responsible for anti-inflammatory repair functions [23]. This stromal cell-initiated immunomodulatory mechanism provides a novel cellular explanation for acupuncture treatment of inflammatory diseases such as *Salmonella* dysentery.

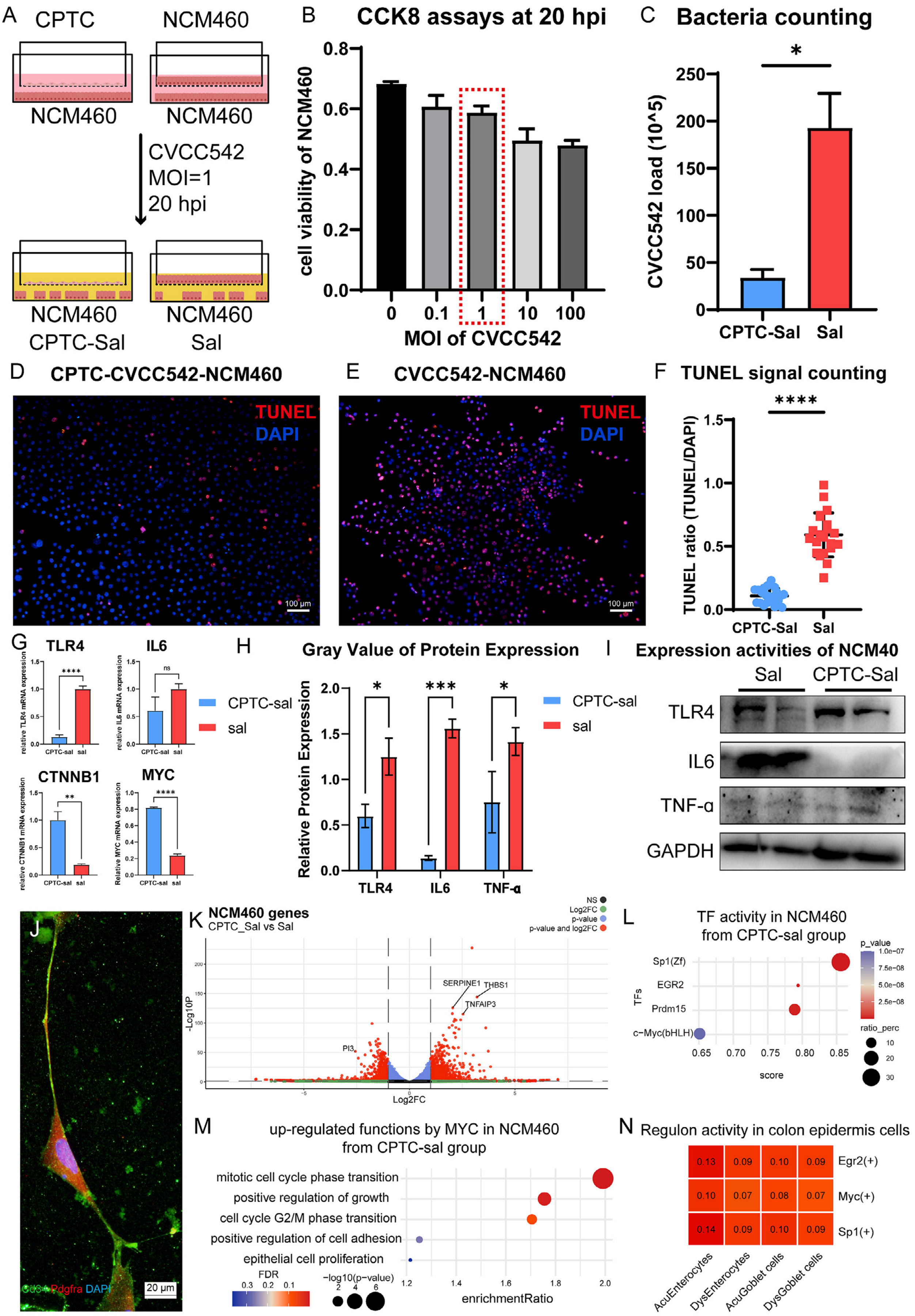

**FIGURE 7** | Legend on next page.

**FIGURE 7** | Fascial CPTC secretome directly rescues epithelial barrier integrity via Wnt/β-catenin/Myc signaling. (A) Schematic of the Transwell non-contact co-culture system. Primary subcutaneous fascia CPTCs are seeded in the upper chamber, NCM460 colonic epithelial cells in the lower chamber, with *Salmonella* (CVCC542) challenge to simulate the dysentery infection microenvironment. (B) CCK8 assay detecting NCM460 cell viability under different multiplicity of infection (MOI), determining MOI = 1 as the optimal challenge dose (red box). (C) Bacterial colony count statistics. Compared to the culture-only group (Sal), the CPTC co-culture group (CPTC-Sal) significantly reduced bacterial load in epithelial cells (*$p < 0.05$). (D–F) Cell apoptosis detection. (D, E) TUNEL (red) and DAPI (blue) dual-staining fluorescence images show the co-culture group (CPTC-CVCC542-NCM460) has significantly reduced DNA fragmentation signals compared to the infection-only group (CVCC542-NCM460). (F) CellProfiler quantification of TUNEL-positive cell proportion (****$p < 0.0001$). (G–I) Expression validation of key inflammation and repair genes. (G) QPCR shows CPTC intervention downregulates TLR4 and IL6, upregulates CTNNB1 and MYC mRNA transcription levels. (H, I) Western Blot and densitometry analysis confirm CPTC significantly suppresses TLR4, IL6, and TNF-α protein expression (*$p < 0.05$,***$p < 0.001$). (J) Typical immunofluorescence morphological characterization of primary subcutaneous fascia CPTCs (green: Cd34; red: Pdgfra; blue: DAPI). (K) NCM460 cell transcriptome volcano plot. Highlighted are significantly upregulated anti-inflammatory gene (TNFAIP3) and downregulated stress gene (PI3) after CPTC intervention. (L, M) Transcriptional regulation analysis. (L) TFHomer transcription factor enrichment analysis shows NCM460 upregulated genes are enriched in c-Myc, EGR2, and other TF binding motifs. (M) GSEA/GO enrichment analysis shows c-Myc-regulated upregulated target genes mainly involve G2/M cell cycle transition and epithelial proliferation functions. (N) SCENIC transcription factor activity analysis of in vivo colonic single-cell data. Heatmap shows Egr2, Myc, and Sp1 Regulon activities in intestinal epithelial cells significantly recover in acupuncture group (Acu) compared to dysentery group (Dys).

Beyond immunomodulation, this study also found that acupuncture intervention could significantly enhance canonical/non-canonical Wnt signaling (such as Wnt2, Wnt5a) from colonic CPTCs to intestinal epithelial cells. In intestinal physiology research, specialized Telocytes located at small intestinal crypt and villus tips (villus tip telocytes, VTT) have been confirmed as essential components of the intestinal stem cell niche. Through secreted Wnt ligands, they maintain basic stem cell functional homeostasis in the intestinal stem cell niche microenvironment, effectively regulating epithelial renewal and repair. As the primary source of intestinal Wnt signaling, depletion of this cell population directly leads to epithelial renewal arrest and intestinal tissue structural collapse [17, 24]. Moreover, recent landmark research has confirmed that Telocytes can even deliver Wnt signals to stem cell membrane surface receptors at extremely short distances through synapse-like contacts, and loss-of-function experiments targeting this synaptic delivery function have demonstrated that synapse-like secretion may be the primary Wnt signal delivery mechanism in the intestinal stem cell microenvironment [25]. Our observation of enhanced Wnt signaling in colonic CPTCs following acupuncture, along with increased proportions of proliferating (G2/M phase) epithelial cells, combined with the high consistency between our identified CPTCs and Telocytes described in international literature in terms of transcriptional features and anatomical location [17], strongly suggests that acupuncture may accelerate re-epithelialization of damaged intestinal mucosa by activating the Wnt secretory function of these "niche cells."

Another key question remains: how do mechanical signals from the body surface cross such vast spatial distances at the microscopic level to reach the visceral colon? Our study provides a compelling answer through the discovery of "transcriptional resonance" between fascial and colonic CPTCs. Despite their vastly different anatomical locations, these two cell types share a highly conserved transcriptional program under both acupuncture and dysentery states (Figure 6D and Figure S7). Background signals characterized by high Wnt concentrations formed in the circulatory system under acupuncture intervention may be the systems biology factor causing this transcriptional resonance. Traditional views hold that Wnt signaling primarily acts in a paracrine manner within stem cell niches. However, our findings challenge this limitation. Based on the "interstitium as an organ" concept proposed by Benias et al. [26], subcutaneous fascia is not merely a supporting structure but a highway for fluid conduction. Our observed elevation of serum Wnt2 suggests that acupuncture may pump locally high-concentration CPTC secretome into circulation through tissue fluid-peripheral circulation exchange via the structurally loose endothelium and lymphatic system [27]. Furthermore, addressing potential skepticism about long-distance transport difficulties for Wnt proteins due to their hydrophobicity, the electron microscopy evidence in this study provides a key answer. EV-associated ultrastructure observed at perivascular fascia (Figure 6B) is consistent with a vesicular transport mechanism for Wnt2; however, EV-mediated long-distance delivery remains one potential mechanism warranting future investigation [28].

More importantly, we need not assume that every Wnt molecule reaching the colon must originate from fascia—virtually all metazoan cells secrete Wnt2, and serum actually serves as a convergence pool for this signal. Elevated serum Wnt2 marks the organism's entry into a pro-regenerative state [13]. In this state, CPTCs of the body surface and viscera are not in a simple "sender-receiver" relationship but rather exist in a state of transcriptional resonance. Wnt signaling in the circulatory system may act as a systemic synchronization signal, lowering the threshold for distant tissues to initiate repair programs [29]. Based on these perspectives, the therapeutic efficacy of acupuncture intervention can be summarized at the systems biology level as follows: a systemic microenvironmental reprogramming characterized primarily by enhanced reparative immunomodulatory signaling.

The transcriptional concordance between fascial and colonic CPTCs ($R = 0.75$, Figure S7D) is fully consistent with the systemic co-regulation model outlined above and does not require a strict linear fascia-to-colon relay. Both CPTC populations could converge on similar transcriptional states by responding in parallel to shared systemic cues—elevated circulating Wnt2, reduced bacterial burden, or acupuncture-associated neuroendocrine output—rather than through serial inter-tissue cargo transfer. This parallel-convergence interpretation is, in fact, the more parsimonious reading of the available evidence: it accommodates the serum Wnt2 elevation (Figure 6C), the convergent transcriptional programs (Figure S7), and the pro-regenerative shift in colonic CPTCs within a single systemic reprogramming model. Directional lineage tracing of Wnt2-loaded vesicles from fascia to circulation to colon remains an important future goal; however, this does not alter the central conclusion that acupuncture establishes a systemic pro-repair milieu in which fascial and colonic CPTCs operate in transcriptional synchrony.

The in vitro co-culture experiments (Figure 7) provide functional evidence consistent with this reparative immunomodulation model. While the cross-species design (rat CPTCs with human NCM460 cells) warrants confirmation in a species-matched system as noted in Limitations, the conserved Wnt/β-catenin pathway (>90% amino acid identity for core components) and the consistent functional endpoints support the interpretation that CPTC-derived signals are sufficient to activate epithelial regenerative programs. We found that merely introducing the subcutaneous fascial CPTC secretome as a microenvironmental component (the secretome signals used by fascial CPTCs and colonic CPTCs to construct microenvironments are highly similar, see Figure 5) was sufficient to reduce *Salmonella* proliferation capacity and inflammatory damage, activate the Wnt/β-catenin/Myc pathway in intestinal epithelial cells, directly reduce infection-induced apoptotic effects, and promote barrier repair-related functions. This finding is transformative; it demonstrates that Wnt signaling—whether locally secreted by colonic CPTCs or delivered from the fascial system—represents a potential direct "survival factor" for colonic epithelial cells. This result aligns perfectly with the latest research demonstrating that peripheral blood Wnt2 is an acute-phase anti-apoptotic factor [13].

Furthermore, traditional perspectives have long held that the therapeutic effects of acupuncture on ulcerative colitis primarily depend on neuro-immune regulatory networks. Previous studies have focused on how acupuncture suppresses pro-inflammatory factor release and regulates macrophage polarization (M1 to M2 transition) by activating the vagus nerve-adrenal axis or cholinergic anti-inflammatory pathway [23, 30, 31]. In these classical models, intestinal epithelial barrier repair is typically viewed as a secondary outcome following inflammation resolution rather than a direct target of acupuncture signaling. Therefore, the in vitro co-culture system in this study also addresses the limitations of this single-perspective research.

We further clarify the evidentiary basis for our conclusions. Convergence across DNN classification, SHAP attribution, and GSEA enrichment provides methodological robustness—demonstrating that findings are not artifacts of a single analytical approach—but does not constitute independent biological replication, as all three methods operate on the same dataset from six animals sharing the same biological noise and animal-level variation. We therefore delineate the following confidence hierarchy: (i) independently validated (high confidence): CPTC secretome rescues epithelial barrier integrity via Wnt/β-catenin (Figure 7, in vitro data independent of scRNA-seq); cross-study G-Index ranking (Table 1, independent animal cohort); (ii) single-cohort transcriptomics (moderate confidence): within-study CPTC ranking, AP-1/Hsp70 mechanotransduction axis, GO:0009612 enrichment; (iii) exploratory (hypothesis-generating): cross-tissue transcriptional correlation and communication network topology.

Limitations: Several limitations should be noted. The in vivo dataset is derived from two biological replicates per group (with a third animal retained as backup for each group); while the core CARSS identification and SHAP-validated AP-1/Hsp70 findings are supported by multi-level evidence (GO:0009612, $p = 0.000839$), inferred communication network topologies and the cross-tissue transcriptional correlation ($R = 0.75$) are best treated as hypothesis-generating pending replication in larger cohorts. The absence of a sham acupuncture control means acupoint specificity cannot be formally separated from general needle-insertion effects, though the selective enrichment of mechanosensory pathways in CPTCs—rather than a generic inflammatory signature—argues for structured mechanical stimulation. Somatic mechanical stimulation activates CPTCs as mechanosensory relay stations. Although prior morphometric studies have documented differential TC distribution and enrichment at classical acupoint sites across multiple species [32, 33], whether the therapeutic effects observed here require acupoint-specific stimulation or extend to generalized mechanical loading of TC-enriched connective tissue remains to be formally established through sham acupuncture controls in future work. The in vitro co-culture pairs rat CPTCs with human NCM460 cells; Wnt pathway components are >90% conserved between species and functional endpoints were consistent, but confirmation in a species-matched system remains desirable. Finally, direct validation of the proposed EV-mediated Wnt2 delivery route—including cargo confirmation, CPTC lineage tracing, and secretion-inhibition experiments—represents a defined priority for future mechanistic work. In summary, this study utilizes the CARSS deep learning framework to redefine the cellular mechanisms of acupuncture. We have delineated a complete chain of events: (1) subcutaneous fascial CPTCs immediately sense and respond to mechanical stress through the AP-1 axis; (2) acupuncture intervention signals are transduced into systemic upregulation of Wnt signaling; (3) subcutaneous fascial CPTCs and colonic CPTCs undergo resonance, while strengthening the Wnt signaling circuit, targeting intestinal epithelium and other intestinal microenvironment cells to mitigate the impact of intestinal inflammatory storms and promote the overall transition of the intestine toward an immune-repair mode. These findings not only validate the scientific basis for acupuncture alleviating bacterial dysentery in distant colonic tissue but also propose serum Wnt2 as a potential quantifiable biomarker for acupuncture intervention efficacy against bacterial dysentery, providing a reusable research paradigm for acupuncture research based on systems biology knowledge.

## Author Contributions

**Haixiang Huang:** conceptualization, methodology, software, formal analysis, investigation, data curation, writing – original draft, visualization. **Zhenwei Zhang:** methodology, formal analysis, investigation, data curation, visualization. **Bingbing Shen:** investigation. **Jianming Yue:** investigation. **Lu Mei:** investigation. **Xudong Zhu:** investigation. **Yonghong Shi:** investigation. **Qianmei Zhu:** investigation. **Yeping Shi:** investigation. **Yifan Luo:** investigation. **Yitong Xing:** investigation. **Meng Dai:** investigation. **Qiusheng Chen:** conceptualization, resources, supervision, project administration, funding acquisition, writing – review and editing.

## Acknowledgments

This study was supported by internal research funds from the authors' institutions.

## Funding

This work was supported by National Natural Science Foundation of China (grant 31872433).

## Conflicts of Interest

The authors declare no conflicts of interest.

## Data Availability Statement

Single-cell RNA sequencing data are available from the CNCB Genome Sequence Archive (GSA) under accession number CRA020447 and the NCBI Gene Expression Omnibus (GEO) database under accession number GSE261489. CARSS analysis scripts are available at https://github.com/NJAU-vet-QSChen-s-lab/Acupuncture-CARSS.

## Supporting Information

Additional supporting information can be found online in the Supporting Information section. **Figure S1:** Architecture and validation of the Cellular Acupuncture Response Scoring System (CARSS). This figure illustrates the deep learning methodology and neural network architecture used in the CARSS framework. The system employs adaptive multilayer perceptron models with configurable parameters including hidden layer sizes (20–128 neurons), network depths (4–8 layers), and activation functions (ReLU, Sigmoid, ELU, Softmax, Hardswish). A two-phase hyperparameter optimization strategy systematically evaluated 21 600 parameter combinations to identify optimal model configurations for each cell type. **Figure S2:** In situ identification and morphological characterization of CPTCs in fascia and colon. Representative immunofluorescence images validating the in situ presence and morphological features of $CD34^+/PDGFR\alpha^+$ CPTCs in both fascia and colon tissues. This provides morphological evidence supporting the single-cell RNA sequencing findings and confirms that these cells genuinely exist in their predicted tissue locations with characteristic telocyte morphology. Scale bar: 40 μm (A, B); 5 μm (B1), 2 μm (C), 20 μm (D). **Figure S3:** Quality control metrics and batch effect correction for scRNA-seq data. Principal component analysis (PCA) and quality control visualizations demonstrating successful batch effect removal and high data quality across samples. This figure shows that cells cluster by biological identity rather than technical batch, confirming the rigor of our single-cell analysis pipeline. **Figure S4:** Resolution of inflammatory signaling and restoration of metabolic homeostasis in colonic effectors following mechanotherapy. SHAP-based functional enrichment and KEGG signaling pathway analysis showing the transition of colonic epithelial cells and lymphocytes from inflammatory/defense mode to metabolic/synthetic mode following acupuncture treatment. This provides molecular phenotypic evidence supporting the functional recovery observed at the tissue level. **Figure S5:** Top 50 SHAP-ranked genes driving DNN model performance across cell types. Heatmap displaying the top 50 genes contributing most to each cell type's DNN model performance, ranked by their SHAP values. Columns represent different cell populations (e.g., faCPTC, faFC, faMC), rows represent SHAP ranking. Color gradient indicates relative feature importance (rank-normalized), with yellow indicating highest contribution. For fascia-derived cells, distinct molecular signatures are identified as key model predictors: (1) Fascia CPTC (faCPTC) features genes related to cellular stress response and repair (such as Hspa1b and Jun) with high contribution. (2) Fascia fibroblasts (faFC) show mixed signatures of inflammatory factors and stress-related genes, with Il6, C7, and Hspa1b ranking high. (3) Fascia myeloid cells (faMC) exhibit strong dependence on immune function and inflammation-related genes, particularly the Ccl chemokine family (e.g., Ccl2, Ccl4). **Figure S6:** High-resolution transcriptomic analysis and functional annotation of colonic cell subpopulations. Bubble plot showing marker expression patterns for 33 cell functional subpopulations in acupuncture and dysentery group colons. By analyzing the specific expression distribution of canonical and functional markers, all unsupervised clustering subpopulations are annotated to reflect each cell type's primary biological function at maximum resolution (e.g., CPTCs_tissue_repair, Macrophage_inflammatory). **Figure S7:** Comparison of Wnt signaling pathway key component expression patterns in subcutaneous fascia and Colonic CPTCs. (A) Heatmap showing average expression levels (*Z*-score normalized) of Wnt signaling pathway core genes in CPTCs from subcutaneous fascia (scFascia) and colon under acupuncture and dysentery models. Red represents high expression, blue represents low expression. (B) Gene density plots in UMAP space displaying transcriptional abundance distribution of 8 key Wnt signaling pathway components (ligands and receptors) in the fused cell population. Darker colors (purple) represent higher gene density. (C) *Concordant Expression Profiles*. Heatmap comparing the expression intensity of identified "Resonance Genes" (*Mfap2, Clec3b, Dbp, Il33, Wnt2, Fgf7, Dapk1*) between colon and fascia tissues, highlighting a conserved transcriptional signature. (D) *Core Resonance Signature*. Scatter plot correlation analysis of the $Log_2$ Fold Change (Acupuncture vs. Dysentery) for key genes in Fascia (X-axis) versus Colon (Y-axis). Scatter plot of $Log_2$ fold change (Acupuncture vs. Dysentery) for the identified resonance genes in fascial CPTCs (x-axis) versus colonic CPTCs (y-axis). A positive correlation was observed ($R = 0.75$, $p = 0.052$), reflecting directional concordance in transcriptional response across both CPTC populations. **Figure S8:** Identification and transcriptional characterization of cell populations in subcutaneous fascia. (A) Uniform Manifold Approximation and Projection (UMAP) visualization of the subcutaneous fascia single-cell landscape. Major cell lineages, including stromal cells (CPTCs, Fibroblasts), immune cells (Macrophages, T cells, B cells), and endothelial cells, are identified and labeled based on their unique transcriptomic profiles. (B–E) Kernel density estimation plots displaying the spatial expression distribution of representative lineage markers in UMAP space: *Pdgfra* (stromal), *Cd3e* (T lymphocytes), *Cd68* (macrophages), and *Aqp1* (endothelial cells). The color gradient indicates gene expression density, with brighter colors (yellow) representing higher density regions. (F) Dot plot illustrating the expression specificity of top marker genes across all annotated cell clusters. The dot size represents the percentage of cells expressing the marker, while the color intensity (red to blue) indicates the average scaled expression level. This profile validates the accuracy of cell type assignment based on canonical marker signatures (e.g., *Cd79a* for B cells, *Cd34*/*Pdgfra* for CPTCs). **Table S1:** Mechanospecific differentially expressed genes in fascial CPTCs following acupuncture. **Table S2:** Cross-study G-index values in forward and reverse transfer directions. **Table S3:** SHAP feature attribution scores for all genes in the cross-study CPTC model. **Table S4:** Dual-evaluation strategy reveals the hierarchy of mechanosensitive cells (within-study evaluation).